%% file: 0-main.tex
\pdfoutput=1 

\documentclass[pdflatex,sn-mathphys-num]{sn-jnl}

\usepackage{graphicx}%
\usepackage{multirow}%
\usepackage{amsmath,amssymb,amsfonts}%
\usepackage{amsthm}%
\usepackage{mathrsfs}%
\usepackage[title]{appendix}%
\usepackage{xcolor}%
\usepackage{textcomp}%
\usepackage{manyfoot}%
\usepackage{booktabs}%
\usepackage{algorithm}%
\usepackage{algorithmicx}%
\usepackage{algpseudocode}%
\usepackage{listings}%
\usepackage{subcaption} 
\RequirePackage{natbib}
\RequirePackage{hyperref}
\usepackage[utf8]{inputenc} 
\usepackage{bbm}
\usepackage{lmodern,babel,adjustbox,booktabs,multirow}
\usepackage{array}  
\usepackage{colortbl}  
\usepackage{rotating}

\theoremstyle{thmstyleone}%
\theoremstyle{thmstyletwo}%

\theoremstyle{thmstylethree}%

\newcommand{\fullname}{\textbf{L}atent \textbf{S}ocial \textbf{D}ynamical-\textbf{S}ystem}
\newcommand{\socialdynamical}{LSDS}

\newcommand{\datasetname}{\textbf{T}witter 20\textbf{20} \textbf{D}ynamical data of \textbf{Pol}itical-related \textbf{A}ccounts}
\newcommand{\dataset}{T20D-PolA}

\DeclareMathAlphabet{\mymathbb}{U}{BOONDOX-ds}{m}{n}

    {\clearpage\pagebreak[4]\global\pdfpageattr\expandafter{\the\pdfpageattr/Rotate 90}}%
    {\clearpage\pagebreak[4]\global\pdfpageattr\expandafter{\the\pdfpageattr/Rotate 0}}%

\begin{document}

\title[Article Title]{A Social Dynamical System for Twitter Analysis}


\author[1]{\fnm{Zhiping} \sur{Xiao}}\email{patricia.xiao@cs.ucla.edu}

\author[2]{\fnm{Xinyu} \sur{Wang}}\email{1800017869@pku.edu.cn}

\author[2]{\fnm{Yifang} \sur{Qin}}\email{qinyifang@pku.edu.cn}

\author[1]{\fnm{Zijie} \sur{Huang}}\email{zijiehjj@gmail.com}

\author*[3,4]{\fnm{Mason A.} \sur{Porter}}\email{mason@math.ucla.edu}

\author*[1]{\fnm{Yizhou} \sur{Sun}}\email{yzsun@cs.ucla.edu}

\affil*[1]{\orgdiv{Department of Computer Science}, \orgname{University of California}, \orgaddress{
\street{580 Portola Plaza}, 
\city{Los Angeles}, 
\postcode{90095}, 
\state{California}, \country{United States of America}}}

\affil[2]{\orgdiv{School of Computer Science}, \orgname{Peking University}, 
\orgaddress{
\street{No. 5 Yiheyuan Road}, 
\city{Beijing}, 
\postcode{100084}, 
\country{People's Republic of China}}}

\affil*[3]{\orgdiv{Department of Mathematics}, \orgname{University of California}, \orgaddress{
\street{520 Portola Plaza}, 
\city{Los Angeles}, 
\postcode{90095}, 
\state{California}, \country{United States of America}}}

\affil*[4]{
\orgname{Santa Fe Institute}, \orgaddress{
\street{1399 Hyde Park Road}, 
\city{Santa Fe}, 
\postcode{87501}, 
\state{New Mexico}, \country{United States of America}}}


\abstract{
Understanding the evolution of public opinion is crucial for informed decision-making in various domains, particularly public affairs.
The rapid growth of social networks, such as Twitter (now rebranded as $\mathbb{X}$), provides an unprecedented opportunity to analyze public opinion at scale without relying on traditional surveys. 
With the rise of deep learning, Graph Neural Networks (GNNs) have shown great promise in modeling online opinion dynamics. Notably, classical opinion dynamics models, such as DeGroot, can be reformulated within a GNN framework.

We introduce {\fullname} (\textbf{\socialdynamical}), a novel framework for modeling the latent dynamics of social media users' opinions based on textual content. 
Since expressed opinions may not fully reflect underlying beliefs, \textbf{\socialdynamical} first encodes post content into latent representations. It then leverages a GraphODE framework, using a GNN-based ODE function to predict future opinions. A decoder subsequently utilizes these predicted latent opinions to perform downstream tasks, such as interaction prediction, which serve as benchmarks for model evaluation.
Our framework is highly flexible, supporting various opinion dynamic models as ODE functions, provided they can be adapted into a GNN-based form. It also accommodates different encoder architectures and is compatible with diverse downstream tasks.

To validate our approach, we constructed dynamic data sets from Twitter data. Experimental results demonstrate the effectiveness of \textbf{\socialdynamical}, highlighting its potential for future applications. 
We plan to publicly release our data set and code upon the publication of this paper.

}




\maketitle

\input{1-introduction.tex}
\input{2-related.tex}
\input{3-definition.tex}
\input{4-model.tex}

\input{5-experiments.tex}
\input{6-conclusion.tex}


\backmatter

\input{7-declarations.tex}






\bibliography{references}

\end{document}

%% file: 1-introduction.tex
\section{Introduction}\label{sec:sds_intro}

Political opinions exert a profound influence on society, shaping public policies, guiding societal norms, and driving collective action~\cite{jost2009political,treier2009nature,kumlin2006learning,kalmoe2020uses,hill2021nastiest}. Furthermore, political ideologies play a pivotal role in determining electoral outcomes~\cite{levitin1979ideological,flamino2023political}.
Public opinion is inherently dynamic, evolving over time in response to social phenomena~\cite{hu2021revealing}. Analyzing these shifts is crucial for informed decision-making across various domains, including policy-making, business, and marketing. For example, changes in public sentiment have contributed to increasing political polarization, a defining characteristic of contemporary politics~\cite{baldassarri2008partisans,pew2017partisan}.

Traditionally, opinion dynamics models--also known as social influence models--have been used to characterize how individuals adjust their opinions in response to their social environments. Notable models include the \textsc{DeGroot} model~\cite{abelson1964mathematical,degroot1974reaching} and the \textsc{Friedkin--Johnsen} (FJ) model~\cite{friedkin1990social}, among others~\cite{hegselmann2002opinion,noorazar2020classical}. These models represent opinions as variables, which may be continuous~\cite{deffuant2002can,hegselmann2002opinion} or discrete~\cite{kozitsin2022general,sznajd2000opinion}, single-dimensional or multi-dimensional~\cite{parsegov2016novel,friedkin2019mathematical,baumann2021emergence}, and evolving over continuous~\cite{abelson1964mathematical,abelson1967mathematical} or discrete time steps~\cite{french1956formal,degroot1974reaching}. When applied to real-world data, these models typically estimate parameters based on mathematical characteristics of the data distributions~\cite{parsegov2016novel} or by experimenting with different parameter settings and selecting the configuration that yields the best empirical performance~\cite{kozitsin2022general}.

Despite their theoretical significance, evaluating social opinion dynamics models requires real-world data sets, as simulated data alone is insufficient for validating their applicability~\cite{cinelli2021echo,de2021no,devia2022framework,kozitsin2022general}. In practice, real-world opinion data sometimes challenge prevailing theoretical assumptions. For instance, an analysis of Reddit data contradicted the widely accepted echo chamber hypothesis~\cite{de2021no}, while studies based on the World Values Survey demonstrated that most existing models exhibit a strong agreement bias, leading to an unrealistic tendency toward consensus~\cite{devia2022framework}.
Online social networks, such as Twitter (now rebranded as $\mathbb{X}$), provide rich sources of dynamic opinion data. Compared to traditional surveys, these platforms offer real-time insights at scale with significantly lower data collection costs. However, their label-free nature necessitates the development of preprocessing techniques for annotating data~\cite{kozitsin2022general,kozitsin2022formal}.

Many existing opinion dynamics models rely on strict mathematical assumptions that limit their flexibility. For example, the framework proposed by Kozitsin~\cite{kozitsin2022general} effectively models opinion updating rules on real-world data sets but imposes strong constraints. One such assumption is that individuals with similar opinions will behave similarly when exposed to comparable external influences. Consequently, the model primarily captures average patterns of opinion formation while assuming equal influence among all agents. This limitation underscores the need for more adaptable and empirically grounded approaches to modeling opinion dynamics.
The traditional methods for tuning the parameters of opinion dynamics models to fit real-world data sets often impose constraints on their expressiveness~\cite{kozitsin2022formal,kozitsin2022general}. In contrast, neural networks (NNs) offer a promising alternative due to their flexibility and ability to optimize parameters efficiently through backpropagation. This eliminates the need for meticulous manual tuning based on theoretical assumptions and empirical observations~\cite{hecht1992theory}.

\begin{figure}
    \centering
    \includegraphics[width=0.8\linewidth]{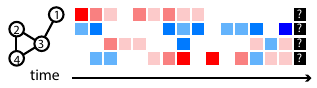}
    \caption{Illustration of opinion dynamics in a network. Nodes represent individuals, and edges denote interactions. Colors indicate observed opinions along a liberal--conservative spectrum, with darker shades representing more extreme opinions. The darkest blue and darkest red correspond to the most extreme positions. White nodes signify instances where opinions are unobserved at the given time.
    }
    \label{fig:socialdynamic_illustration}
\end{figure}

With the rise of deep learning, Graph Neural Networks (GNNs) have become widely used in social network analysis~\cite{sharma2024survey,ju2024comprehensive}. Social networks can be naturally represented as node-edge graphs, where nodes correspond to accounts and edges denote follower–followee relationships. As illustrated in Figure~\ref{fig:socialdynamic_illustration}, we frame the prediction of individuals' future opinions as forecasting future values on nodes while learning the underlying updating rules governing these values.
Existing GNN models predominantly focus on static social networks~\cite{10.1145/3394486.3403275} or analyze temporal graphs by examining discrete snapshots~\cite{min2021stgsn}. Some models address continuous-time dynamics but rely on strict assumptions, limiting their applicability to real-world scenarios~\cite{skarding2021foundations}. Meanwhile, in fields such as physics, Graph-based Neural Ordinary Differential Equations (GraphODE) have demonstrated success in modeling dynamic multi-agent systems~\cite{NEURIPS2020_ba484941}, suggesting their strong potential for learning opinion-update rules in social networks.

We introduce {\fullname} (\textbf{\socialdynamical}), a novel framework for modeling opinion dynamics in multi-agent social network systems. {\socialdynamical} learns the updating rules of latent opinions through three key components: an encoder, a GraphODE module, and a decoder.
Since ground-truth labels for opinions are unavailable, we estimate opinions from textual data. We leverage a pre-trained large language model (LLM)~\cite{reimers2019sentence} to transform posts into vector-space representations. {\socialdynamical} then employs an encoder to map these representations into latent opinions while incorporating social interactions (e.g., mentions, retweets, likes, replies). This design accounts for the discrepancy between expressed and underlying opinions.
Next, a GraphODE framework models the temporal evolution of these latent opinions. Using the encoder’s outputs as initial conditions, a GraphODE solver employs a GNN-based ODE function to predict future opinions. 
Finally, a decoder extracts meaningful outcomes, such as polarization scores or the likelihood of future interactions (e.g., retweeting behavior), which serve as both downstream tasks and evaluation benchmarks.

Experimental results confirm the effectiveness of {\socialdynamical}. Moreover, the framework is highly flexible, supporting integration with various opinion dynamics models, adaptable GNN-based GraphODE functions, diverse encoder architectures, and a broad range of decoder tasks.

Our paper makes the following contributions:
\begin{enumerate}
    \item[(1)] We propose a novel {\fullname} framework~(\textbf{\socialdynamical}) that models opinion dynamics in real-world online social networks without requiring manual labeling by human experts. The framework effectively captures discrepancies between observed and latent opinions while offering flexibility to integrate various opinion dynamics models, diverse graph encoders, and a broad range of downstream tasks.
    \item[(2)] To facilitate further research, we introduce the {\datasetname} (\textbf{\dataset}) data set, which contains observations from Twitter accounts throughout 2020 (see Section~\ref{subsec::sds_dataset}). This data set provides time-dependent textual and interaction data, making it a valuable resource for studying opinion dynamics in online social networks.
    \item[(3)] Our experiments yield key insights. Notably, the choice of an effective encoder, which accurately captures the initial state of opinions, plays a more critical role than the specific updating rules. Additionally, we observe that social media users rarely switch between liberal and conservative ideological stances, suggesting that political opinions on social networks exhibit significant stability over time.
\end{enumerate}

%% file: 2-related.tex
\section{Related Work and Preliminary Discussions}\label{sec:sds_related}

In this section, we review related works from four key perspectives: opinion dynamics models, graph neural networks, neural ordinary differential equations, and variational encoders.

\subsection{Models of Opinion Dynamics}

Social scientists have long been interested in modeling how opinions evolve through interactions~\cite{lorenz2007continuous,das2014modeling,noorazar2020recent,brooks2022emergence,de2021no}.
Various opinion formation models have been developed as mathematical frameworks to simulate and predict opinion dynamics over time.
These models include the \textsc{DeGroot} model~\cite{abelson1964mathematical,degroot1974reaching}, the \textsc{Friedkin--Johnsen} (FJ) model~\cite{friedkin1990social}, the \textsc{Hegselmann--Krause} (HK) model which is a type of bounded-confidence model (BCM)~\cite{rainer2002opinion}, and many others~\cite{hegselmann2002opinion,noorazar2020classical}.
These models represent opinions using either continuous~\cite{deffuant2002can,hegselmann2002opinion} or discrete~\cite{kozitsin2022general,sznajd2000opinion} variables, which may exist in single-dimensional or multi-dimensional spaces~\cite{parsegov2016novel,friedkin2019mathematical,baumann2021emergence}. Moreover, they can be formulated in either continuous-time~\cite{abelson1964mathematical,abelson1967mathematical} or discrete-time~\cite{french1956formal,degroot1974reaching} settings.
These research efforts are becoming increasingly significant for human society, particularly in the age of the internet, where social networks have gained widespread popularity~\cite{noorazar2020classical,de2021no}.

While many opinion dynamics models were initially developed using simulated data, recent efforts have sought to bridge theory and empirical observations by applying these models to real-world data sets. Some studies evaluate these models using survey data~\cite{devia2022framework}, while others rely on social network data~\cite{kozitsin2022general,cinelli2021echo,kozitsin2022formal}. 
Studies evaluating opinion dynamics models on real-world data, such as those using survey data, suggest that these models often exhibit an unrealistic bias toward consensus~\cite{devia2022framework}. 
However, accurately assessing their predictive power in real-world opinion-updating mechanisms remains challenging, as real-world data sets often violate the assumptions used in simulated data.
 
While online social networks provide access to vast amounts of data compared to traditional methods like surveys, evaluating opinion dynamics models on these data sets is significantly more challenging due to the absence of ground-truth labels.
Some studies circumvent this issue by leveraging pre-existing models~\cite{kozitsin2020modeling} to infer opinion labels at scale, treating these inferred labels as ground truth~\cite{kozitsin2022general,kozitsin2022formal}.
In many studies that applied opinion dynamics models to real-world data sets, researchers estimated model parameters (e.g., weight matrices) from observed data by extending existing algorithms. These parameter estimation methods typically rely on mathematical properties derived from data distributions and are rigorously justified through analytical reasoning~\cite{parsegov2016novel}.

Recently, the Sociologically-Informed Neural Network (SINN) framework~\cite{okawa2022predicting} demonstrated the potential of neural networks in modeling opinion dynamics. This approach used neural architectures to learn the derivatives of opinion variables in a discrete-time dynamical system. The authors first employed a predefined opinion model to establish ground-truth opinion-update rules and then trained a neural network to approximate these rules. Their findings highlight the capacity of neural networks to replicate traditional opinion models while offering greater flexibility. Moreover, their results suggest that the parameters governing opinion dynamics can be effectively represented within a neural-network framework.
Sociologically-Informed Graph Neural Network (SIGNN) followed this line of work~\cite{yang2025sociologically} and further validated that the Graph Neural Network (GNN) architecture is well-suited for modeling opinion dynamics in social networks. Their findings demonstrate that GNNs outperform SINN, which relies solely on traditional neural networks (NNs) rather than leveraging graph-based structures.

Building upon these insights, we propose \textbf{\socialdynamical}, a flexible framework that integrates opinion dynamics models with more expressive, data-driven approaches, such as neural networks. \textbf{\socialdynamical} enables direct learning of opinion-update rules from real-world data sets and facilitates empirical validation against observed opinion shifts. This framework represents a step toward more robust and adaptable models of opinion evolution, bridging the gap between classical opinion dynamics and modern machine learning techniques.

\subsection{Graph Neural Networks}

Graph neural networks (GNNs) are neural network models designed for graph-structured data, consisting of nodes and their connecting edges~\cite{ju2024comprehensive}. By leveraging neural networks, GNNs effectively address graph-based problems and have demonstrated success across various fields, attracting significant attention~\cite{zhou2020graph,wu2020comprehensive,zhang2020deep,abadal2021computing,rahmani2023graph}.

GNNs are broadly categorized based on their architecture and the tasks they target, including recurrent-based GNNs, convolution-based GNNs, spatiotemporal GNNs, graph autoencoders (GAEs), graph adversarial networks, and graph reinforcement-learning models~\cite{rahmani2023graph}. These categories often overlap, meaning a single GNN model may belong to multiple classes. 
This work focuses specifically on convolution-based GNNs.

Graph convolutions~\cite{balcilar2020bridging} operate in either the spectral domain~\cite{henaff2015deep} or the spatial domain~\cite{zhu2018modelling}. Spectral convolutional GNNs leverage spectral graph theory, requiring access to the entire graph, while spatial convolutional GNNs consider only a node’s local neighborhood. For example, GCN~\cite{kipf2016semi} is a spectral model that normalizes the adjacency matrix using the symmetrically normalized Laplacian of the entire graph. In contrast, models such as \textsc{NRI}~\cite{kipf2018neural}, GIN~\cite{xu2019powerful}, \textsc{GraphSAGE}~\cite{hamilton2017inductive}, and many other convolutional GNNs, are spatial models~\cite{balcilar2021analyzing,zhang2022graph}.
While spectral convolutional GNNs offer provable theoretical properties~\cite{fu2022p,zhu2022generalized,ruiz2023transferability}, spatial convolutional GNNs are generally more computationally efficient and cost-effective to train.
Spatiotemporal GNNs integrate GNN components with recurrent neural networks (RNNs) to model sequential data. These models have been particularly effective in applications such as traffic forecasting~\cite{zhao2019tgcn,guan2022dynagraph}.

Despite their widespread adoption and demonstrated success, GNNs remain underutilized in analyzing dynamic graph data~\cite{skarding2021foundations}. Investigating their potential for studying evolving real-world networks, such as opinion dynamics in social networks, presents an open and challenging research direction.

\subsection{Neural Ordinary Differential Equations}

Solving ordinary differential equations (ODEs) is a well-established field that researchers have explored for centuries~\cite{ince1956ordinary,hartman2002ordinary,miller2014ordinary}. In mathematics, the term ``ordinary'' distinguishes ODEs from partial differential equations (PDEs) by indicating that only a single independent variable is involved.
For example, in an ODE with dependent variable $y$ and independent variable $x$, the only derivative term present is $\frac{dy}{dx}$. 
In contrast, a PDE involves multiple independent variables, such as $x$ and $w$, can contain terms like $\frac{\partial y}{\partial x}$ and $\frac{\partial y}{\partial w}$.
A general ODE takes the form
\begin{equation}\label{eq:ode_relatedwork}
    \frac{dy}{dx} = f(x, y)\,,
\end{equation}
where $x$ is the independent variable, and $y$ is the dependent variable. We can also denote the $y$ value with a given $x_i$ value by $y(x_i)$.
The function $f(x, y)$ governs the instantaneous rate of change of $y$ with respect to $x$.
Additionally, The ODE~(\ref{eq:ode_relatedwork}) is subject to an initial condition $y(x_0) = y_0$ or a boundary condition $y(x_a) = y_a$, where $y_0$ or $y_a$ are known values. These conditions are essential for determining a unique solution among all possible solutions.
In an ODE, both time and space are continuous. Various numerical methods exist for solving ODEs~\cite{muqri2021study}.

The \textsc{Neural-ODE} model~\cite{chen2018neural} integrates a neural network with an ODE solver by using the network to represent the ODE function $f(x,y)$. This approach extends neural networks from discrete to continuous spaces, enabling them to approximate derivatives continuously.
The contributions of the \textsc{Neural-ODE} model are as follows:
\begin{enumerate}
    \item \textsc{Neural-ODE} re-implements traditional ODE solvers (e.g., \textsc{Runge--Kutta}) to be optimized in an end-to-end manner, making them compatible with deep learning.
    \item By replacing algebraic formulas with neural networks for defining ODE functions, \textsc{Neural-ODE} enhances the expressiveness and flexibility of ODE models, making them suitable for systems where the exact ODE formulation is unknown.
\end{enumerate}
For example, ODE solvers facilitate the construction of continuous versions of neural-network layer-wise gradient updates. Additionally, neural ODEs can model the evolution of node representations in continuous time. One notable example is the \textsc{Latent Graph ODE (LG-ODE)} model~\cite{NEURIPS2020_ba484941}, which successfully analyzed time-dependent physical system graphs, such as predicting the movement of bouncing balls.
Our work follows this research direction; however, we focus specifically on modeling opinion dynamics in social networks.

\subsection{Variational Autoencoder}

A variational autoencoder (VAE)~\cite{kingma2013auto} is a neural-network framework that applies variational inference for training. Structured as an autoencoder, it remains differentiable, enabling optimization via backpropagation. The VAE consists of two key components~\cite{ghojogh2021factor}:
\begin{itemize}
    \item Encoder: Maps the input data to a latent-space distribution.
    \item Decoder: Samples from this latent distribution to generate a hidden representation and reconstruct the original input.
\end{itemize}
VAEs are widely used in models that track changes in node states within a network, such as \textsc{NRI}~\cite{kipf2018neural} and \textsc{LG-ODE}~\cite{NEURIPS2020_ba484941}). These models leverage the VAE framework for training.
We adopt a similar training framework as a standard, unsupervised VAE. However, since our data set contains partial labels, we adopt a semi-supervised training pipeline rather than a strictly unsupervised VAE.
While we retain the encoder and latent-space representation, our decoder is designed to learn many other objectives beyond input reconstruction.

%% file: 3-definition.tex
\section{Problem Definition}\label{sec:sds_definition}

We aim to model how opinions evolve within social networks by capturing the interactions between accounts over time. In this section, we formally define the dynamic social graph and describe the signals used to estimate accounts' opinions.

\subsection{Dynamic Social Graph}

Social networks are commonly represented as node-edge graphs, where nodes correspond to accounts and edges denote follower-followee relationships. Dynamic graphs can vary in structure--both nodes and edges may be time-dependent or static~\cite{skarding2021foundations}.
We model our social dynamical system as a multi-agent dynamical system, where each account’s opinion evolves over time. Given that follower--followee relationships change infrequently, we treat these connections as static edges. Instead, we focus on modeling the temporal evolution of opinions at each node. Accordingly, we study dynamical processes on a time-independent network structure~\cite{porter2016dynamical}, an approach commonly used in opinion dynamics models applied to real-world social networks~\cite{kozitsin2022general}.

We define a social network as a graph $\mathcal{G} = (\mathcal{V}, \mathcal{E})$, where:
\begin{itemize}
    \item $\mathcal{V} = { v_1, v_2, \dots, v_N }$ is the set of $N$ nodes (accounts).
    \item $\mathcal{E}$ is the set of edges representing follower--followee relationships, where $\langle v_i, v_j \rangle \in \mathcal{E}$ indicates that account $v_i$ follows account $v_j$.
    \item The graph $\mathcal{G}$ has $M = |\mathcal{E}|$ edges.
\end{itemize}

We infer accounts' opinions from the textual content of their posts. Each account $v_i$ has a series of observations: $o_i = \{ \mathbf{o}_i^t \}$. 
For any account $v_i$ and any time $t$, the observation $\mathbf{o}_i^t \in \mathbb{R}^D$ is the vector-space representation of the corresponding text content at time $t$.
In practice, we generate $\mathbf{o}_i^t$ from the account's raw tweet text using the pretrained \textsc{Sentence-BERT} model~\cite{reimers2019sentence}. If an account posts multiple tweets at time $t$, we take the mean embedding of all tweets to obtain a single observation $\mathbf{o}_i^t$.

As each account posts at an individual frequency, every account $v_i$ has a unique observation time sequence $\{ t_i^j \}_{j=0}^{T_i}$.
Let $\mathbf{z}_i^t$ denote the hidden representation of account $v_i$ at time $t$; the quantity $\mathbf{z}_i^t$ is account $v_i$'s latent opinion. Our goal is to model how $\mathbf{z}_i^t$ evolve according to time $t$.

\subsection{Opinion Evaluation}\label{sec:opinion_eval}

We evaluate our model’s ability to predict future opinions using four distinct tasks: interaction prediction, polarity prediction, polarity classification, and text-embedding reconstruction. Each task serves as a different decoder objective.

\subsubsection{Interaction Prediction}

To assess the quality of latent opinion representations $\mathbf{z}_i^t$ and $\mathbf{z}_j^t$, we use them to infer interactions between accounts over time.

While our underlying network structure is based on static follow relationships, we model time-dependent interactions such as mentions, retweets, replies, and likes as dynamic edges: $\{ \mathcal{E}_{\mathrm{mention}}, \mathcal{E}_{\mathrm{retweet}}, \mathcal{E}_{\mathrm{reply}}, \mathcal{E}_{\mathrm{like}}\}$, where $\langle i, j, t\rangle \in \mathcal{E}_{\mathrm{mention}}$ indicates that account $v_i$ mentioned account $v_j$ in a post at time $t$.

To evaluate how well our model captures opinion-updating rules, we compute an edge-prediction score:
\begin{displaymath}
    \mathrm{Score}(i, j, t, \mathrm{R}) = f_{\mathrm{R}}(\mathbf{z}_i^t, \mathbf{z}_j^t)\,,
\end{displaymath}
where $f_{\mathrm{R}}$ (for $\mathrm{R} \in \{ \mathrm{mention}, \mathrm{retweet}, \mathrm{reply}, \mathrm{like} \}$) measures the likelihood of an interaction between accounts $v_i$ and $v_j$ based on their latent opinions at time $t$.

\subsubsection{Polarity Prediction}

A major challenge in analyzing real-world social network data is the lack of ground-truth labels. Since our data set contains political discourse, we infer political polarity labels for tweets.

Using the pretrained polarity-aware embedding model (PEM)~\cite{xiao2023detecting}, which assigns a scalar polarity score to each tweet, we estimate the political polarity $p_i^t$ of an account’s tweet and aim to recover this value from the hidden representation $\mathbf{z}_i^t$. The more accurate the reconstruction, the stronger the model’s ability to capture opinion dynamics.

\subsubsection{Polarity Classification}

Polarity classification is similar to polarity prediction, but instead of estimating a continuous polarity score, we predict a binary classification label (liberal vs. conservative).

Using the PEM model~\cite{xiao2023detecting}, we first determine the tweet's political alignment and then train our model to classify the latent opinion $\mathbf{z}_i^t$ into one of two categories.

\subsubsection{Text-Embedding Reconstruction}

In an unsupervised setting, where no labeled training data is available, we evaluate our model by reconstructing future tweet embeddings.

Given an account’s ground-truth future tweet embedding $\mathbf{o}_i^t$, we train a decoder to reconstruct it from the latent representation $\mathbf{z}_i^t$:
\begin{displaymath}
    \Tilde{\mathbf{o}}_i^t = g(\mathbf{z}_i^t) \,,
\end{displaymath}
where $g(\cdot)$ is a decoder function. The closer $\Tilde{\mathbf{o}}_i^t$ is to $\mathbf{o}_i^t$, the better our model captures social opinion dynamics.

%% file: 4-model.tex
\section{Methodology}\label{sec:sds_model}

\begin{figure}[ht]
    \centering
    \includegraphics[width=1.0\columnwidth]{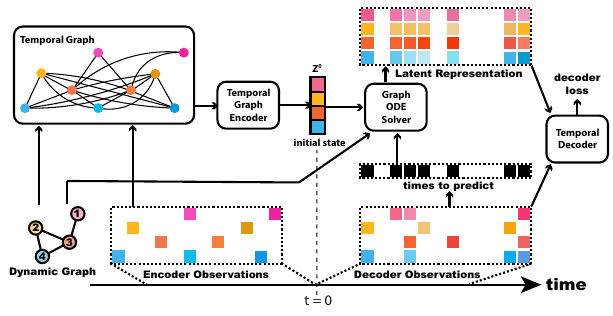}
    \caption{A schematic overview of our model's architecture. We implement the graph ODE function $g_i$ as a GNN function, where all edges in the network depicts the follower--followee relationships in the original graph (i.e., the graph on the bottom-left corner of this Figure).}
    \label{fig:social_dynamical_overview}
\end{figure}

Our \textbf{\socialdynamical} model consists of three main components: a time-dependent encoder, an ODE solver that handles a GNN ODE function, and a time-dependent decoder. 
In Figure~\ref{fig:social_dynamical_overview}, we give a schematic overview of how these components work together in our model.

The encoder transforms the observations of accounts into the initial state of their latent opinion representations. It is time-dependent because it considers historical observations before the starting time $t=0$ by using both observed content and the time stamps. It aims to encode the hidden space of the initial state $\mathbf{z}_i^0$ from all observations $\mathbf{o}_i^{t'}$ before the starting time (i.e., for all observations with $t' < 0$). 
To achieve this goal, we first construct a {temporal graph} $\mathcal{G}_\mathrm{temporal}$ from these observations. 
In this {temporal graph}, each node represents an observation.  
For example, in Figure~\ref{fig:social_dynamical_overview}, we show how our framework works with an example of $4$ accounts (social network is shown on the bottom left corner, and we denote the accounts' nodes as node 1, 2, 3 and 4). There are $2$ observations on {node 1} before $t=0$ and $2$ observations on {node 3}. Therefore, {node 1} in the social network results in $2$ nodes $\{ o_{1a}, o_{1b} \}$ in the {temporal graph}, and similarly {node 3} corresponds to $2$ other {temporal-graph} nodes $\{ o_{3a}, o_{3b} \}$.
An edge in a {temporal graph} represents a follower--followee relationship between the corresponding accounts.
For example, in Figure~\ref{fig:social_dynamical_overview}, {node 1} is adjacent to {node 3}. The {temporal graph} then has the edges: $\{ (o_{1a}, o_{3a}), (o_{1a}, o_{3b}), (o_{1b}, o_{3a}), (o_{1b}, o_{3b}) \}$.
After constructing a {temporal graph}, we encode it with a {temporal-graph encoder}, whose output will allow us to generate the initial latent states $\mathbf{z}_i^0$ of all accounts ($i = 1, 2, \dots, N$).
Ideally, the latent representation $\mathbf{z}_i^t$ captures the latent opinions of account $v_i$ at time $t$. This representation is related to the observation $o_i$ but is not necessarily identical to the direct observation.

An ODE solver allows us to numerically integrate an ODE with a GNN ODE function on the right-hand side. 
The GNN is designed as a message-passing framework, so traditional opinion models such as \textsc{DeGroot} can also fit into this framework (see Section~\ref{subsec:ode_solver}). It aims to provide $\mathbf{z}_i^t$ from the initial state $\mathbf{z}_i^0$, the target time $t \in \mathbb{R}$, and the graph-ODE function.

The decoder uses a hidden representation such as $\mathbf{z}_i^t$ to do downstream tasks. We can evaluate our model's effectiveness from its performance on the downstream tasks.
The decoder is also time-dependent because the downstream tasks use time-dependent hidden representations as their inputs.

\subsection{The Time-Dependent Encoder}

In this subsection, we introduce the architecture of our {temporal-graph encoder}, and another variant of time-dependent encoder that also works in our framework, serving as a baseline encoder (see Section~\ref{subsubsec:gcn_baseline}).

\subsubsection{The Temporal Graph Encoder}\label{sec:encoder}

We use a {temporal-graph encoder} to encode the observation sequences $o_i = \{ \mathbf{o}_i^{t'} | t' < 0 \}$ before the starting time $t=0$ for each account $v_i$ (with $i \in \{ 1, 2, \dots, N\}$) into a factorized distribution of initial hidden states:
\begin{equation}
    q_\phi(\mathbf{Z}^0) = \prod_{i=1}^N q_\phi(\mathbf{z}_i^0 | o_1, o_2, \dots, o_N )\,.
\end{equation}
Because this multi-agent dynamical system consists of many coupled components, similar to the design of \textsc{LG-ODE}~\cite{NEURIPS2020_ba484941}, we choose to use a model that captures interactions between the nodes, instead of simply modeling each node's observation sequence separately using RNNs~\cite{rubanova2019latent}.

We use a similar encoder architecture as in the \textsc{LG-ODE} model. Our encoder has two phases: (1) a dynamic-node-representation learning (DNRL) phase, in which we learn a function $f_\mathrm{embed}$ that learns a structural contextualized embedding $\mathbf{h}_i^t$ for each observation $\mathbf{o}_i^t$ of node $i$ at time $t$; (2) a temporal self-attention (TSA) phase, in which we learn a function $f_\mathrm{readout}$ to represent the sequence $\{ \mathbf{h}_i^t | t < 0 \}$ of each account $v_i$ as a fixed-length vector $\mathbf{u}_i$, which we then use to determine the posterior distribution of the initial state $\mathbf{Z}^0$ (i.e., the matrix who stacks $N$ vectors $\{\mathbf{z}_i^0 | i = 1, 2, \dots, N\}$ together).

In the DNRL phase, we learn the function $f_\mathrm{embed}$, which is a spatiotemporal GNN model. In this model, we first construct a {temporal graph} $\mathcal{G}_\mathrm{temporal} = (O, E)$, where each node in $O$ represents an observation $\mathbf{o}_i^t$. An edge exists between the nodes representing $\mathbf{o}_i^{t_1}$ and $\mathbf{o}_j^{t_2}$ if and only if node $v_i$ and $v_j$ are adjacent in the input graph $\mathcal{G}$. We also consider the impact of previous observations of the same account, so nodes with observations $\mathbf{o}_i^{t_1}$ and $\mathbf{o}_j^{t_2}$ are adjacent when $i=j$.

For each layer $l$, we represent the hidden representation of $o_\mathrm{src} \in O$ as $\mathbf{h}_\mathrm{src}^{(l)}$ and $o_\mathrm{tgt} \in O$ as $\mathbf{h}_\mathrm{tgt}^{(l)}$, respectively. The nodes learn from their neighborhood via a standard message-passing and aggregation paradigm~\cite{he2023message}, where we use attended message-passing and use a summation operation as its aggregation function. That is,
\begin{equation}
    \mathbf{h}_\mathrm{tgt}^{(l)} = \mathbf{h}_\mathrm{tgt}^{(l-1)} + \sigma\Big(\sum_{\mathrm{src} \in \mathcal{N}_{\mathrm{tgt}}} \big( \mathrm{Attention}(\mathbf{h}_\mathrm{src}^{(l-1)}, \mathbf{h}_\mathrm{tgt}^{(l-1)}) \cdot \mathrm{Message}(\mathbf{h}_\mathrm{src}^{(l-1)}, \Delta t(\mathrm{src}, \mathrm{tgt}) )  \big) \Big)\,,
\end{equation}
where $\mathcal{N}_{\mathrm{tgt}}$ is the neighborhood of node $o_\mathrm{tgt}$ (i.e., all nodes that are adjacent to $o_\mathrm{tgt}$). We assume that any source node $o_\mathrm{src}$ is adjacent to the target node $o_\mathrm{tgt}$, denoted as $\mathrm{src} \in \mathcal{N}_{\mathrm{tgt}}$. The quantity $\Delta t(\mathrm{src}, \mathrm{tgt})$ is the temporal gap between the two observations $\mathrm{src}$ and $\mathrm{tgt}$. 
The time-dependent message function $\mathrm{Message}(\cdot)$ is 
\begin{equation}
\begin{aligned}
    & \mathrm{Message}( \mathbf{h}_\mathrm{src}^{(l-1)}, \Delta t(\mathrm{src}, \mathrm{tgt}) ) = \mathbf{W}_\mathrm{val} \hat{\mathbf{h}}_\mathrm{src}^{(l-1)}\,, \\
    \text{where } & \hat{\mathbf{h}}_\mathrm{src}^{(l-1)} = \sigma(\mathbf{W}_\mathrm{temp}[ \mathbf{h}_\mathrm{src}^{(l-1)} \mathbin\Vert \Delta t(\mathrm{src}, \mathrm{tgt}) ]) + \mathrm{TE}( \Delta t(\mathrm{src}, \mathrm{tgt}) ) \,,
\end{aligned}
\end{equation}
the linear transformations $\mathbf{W}_\mathrm{temp}$ and $\mathbf{W}_\mathrm{val}$ ensure that the output dimension is the same as the input dimension, $\mathrm{TE}$ is a projection function that projects the temporal observation gap $\Delta t(\mathrm{src}, \mathrm{tgt})$ onto a vector with the same dimensionality as the hidden space (i.e., the same dimensionality as $\mathbf{h}_\mathrm{src}^{(l-1)}$). We separately project its odd and even dimensions:
\begin{displaymath}
\begin{aligned}
    \mathrm{TE}(\Delta t)_{2 i + 1} = \cos(\Delta t / 10000^{2 i / d})\,, \quad \mathrm{TE}(\Delta t)_{2 i} = \sin(\Delta t / 10000^{2 i / d})\,.
\end{aligned}
\end{displaymath}
The attention function $\mathrm{Attention}(\cdot)$ is based on a self-attention mechanism~\cite{vaswani2017attention}. We introduce the linear transformations $\mathbf{W}_\mathrm{key}$ and $\mathbf{W}_\mathrm{que}$, which have the same dimensionality as $\mathbf{W}_\mathrm{val}$. 
The transformations $\mathbf{W}_\mathrm{key}$, $\mathbf{W}_\mathrm{que}$, and $\mathbf{W}_\mathrm{val}$ together project the input node representations $\hat{\mathbf{h}}_\mathrm{src}^{(l-1)}$ into values $\mathbf{W}_\mathrm{val} \hat{\mathbf{h}}_\mathrm{src}^{(l-1)}$, keys $\mathbf{W}_\mathrm{key} \hat{\mathbf{h}}_\mathrm{src}^{(l-1)}$, and queries $\mathbf{W}_\mathrm{que} \mathbf{h}_\mathrm{tgt}^{(l-1)}$. We then define the attention function $\mathrm{Attention}(\cdot)$ as
\begin{equation}
    \mathrm{Attention}(\mathbf{h}_\mathrm{src}^{(l-1)}, \mathbf{h}_\mathrm{tgt}^{(l-1)}) = (\mathbf{W}_\mathrm{key} \hat{\mathbf{h}}_\mathrm{src}^{(l-1)})^T (\mathbf{W}_\mathrm{que} \mathbf{h}_\mathrm{tgt}^{(l-1)}) \cdot \frac{1}{\sqrt{d}} \,,
\end{equation}
where $d$ is the hidden dimensionality.
Finally, we stack the $L = 2$ layers of the spatiotemporal convolutional layers to obtain the representation $\mathbf{h}_i^t = \mathbf{h}_{\mathrm{tgt}}^{(2)}$, where the corresponding node with observation $\mathbf{o}_i^t$ in $\mathcal{G}_\mathrm{temporal} = (O, E)$ is $o_\mathrm{tgt}$.

We have a spatiotemporal embedding $\mathbf{h}_i^t$ of each $\mathbf{o}_i^t$. However, each individual account $v_i$ can have an observation sequence $o_i$ of variable length. To encode the sequence as a fixed-dimensional vector $\mathbf{u}_i$ and then derive the hidden-representation distribution's mean and standard deviation $\boldsymbol\mu_i, \boldsymbol\sigma_i \in \mathbb{R}^d$ from it, we use the TSA phase. In the TSA phase, we learn $f_\mathrm{readout}$. That is,
\begin{displaymath}
    [\boldsymbol\mu_i \mathbin\Vert \boldsymbol\sigma_i]  = f_\mathrm{readout}(\{ \mathbf{h}_i^t | t \in \{ t_i^j \}_{j=0}^{T_i}, t < 0 \})\,,
\end{displaymath}
where $f_\mathrm{readout}$ outputs a matrix $[\boldsymbol\mu_i \mathbin\Vert \boldsymbol\sigma_i]$, which denotes the concatenation of $\boldsymbol\mu_i$ and $\boldsymbol\sigma_i$. We use $[\cdot \mathbin\Vert \cdot]$ to represent the concatenation operation.

We incorporate temporal information into the embedding $\mathbf{h}_i^t$ and write
\begin{displaymath}
    \hat{\mathbf{h}}_i^t = \sigma(\mathbf{W}_\mathrm{temp} [ \mathbf{h}_i^t \mathbin\Vert \Delta t ] ) + \mathrm{TE}(\Delta t)\,,
\end{displaymath}
where $\Delta t = t - t_\mathrm{start}$ is the difference between the current observation and the earliest observation. This step is very similar to how we compute $\mathrm{Message}(\cdot)$.

We compute a global sequence vector $\mathbf{a}_i \in \mathbb{R}^{d}$ and a linear transformation $\mathbf{W}_a \in \mathbb{R}^{d \times d}$ by calculating
\begin{displaymath}
    \mathbf{a}_i = \tanh\left( \left(\frac{1}{\left|\left\{o_i | t < 0\right\}\right|} \sum_{t} \hat{\mathbf{h}}_i^t\right) \mathbf{W}_a \right)\,, \qquad \mathbf{u}_i = \frac{1}{\left|\left\{o_i | t < 0\right\}\right|} \sum_t \sigma(\mathbf{a}_i^T \hat{\mathbf{h}}_i^t) \hat{\mathbf{h}}_i^t\,,
\end{displaymath}
where $|\{o_i | t < 0\}|$ is the observed sequence length that is accessible by the encoder. We then have another transformation layer:
\begin{equation}\label{eq:encoder_posterior}
    [\boldsymbol\mu_i \mathbin\Vert \boldsymbol\sigma_i] = f_\mathrm{posterior}(\mathbf{u}_i)\,.
\end{equation}
In practice, we implement $f_\mathrm{posterior}$ as a linear-transformation layer with input dimension $d$ and output dimension $2d$.

With our approximate posterior distribution and our hidden-space distribution parameters $\boldsymbol\mu_i$ and $\boldsymbol\sigma_i$, we can sample the initial state $\mathbf{z}_i^0$ from the normal distribution with mean $\boldsymbol\mu_i$ and standard deviation $\boldsymbol\sigma_i$. That is,
\begin{equation}
    \mathbf{z}_i^0 \sim q_\phi(\mathbf{z}_i^0 | o_1, o_2, \dots, o_N) = \mathcal{N}(\boldsymbol\mu_i, \boldsymbol\sigma_i)\,.
\end{equation}

\subsubsection{GCN Encoder with Time-Dependent Input Features}\label{subsubsec:gcn_baseline}

Another option of our framework's time-dependent encoder component is a standard \textsc{Graph Convolutional Network (GCN)}~\cite{kipf2016semi} encoder. This encoder has the identical GNN model architecture as in the original GCN paper and code.\footnote{See \url{https://github.com/tkipf/pygcn} for GCN code.} This version of the encoder contains a $2$-layer graph-convolutional architecture, with the layer-wise update rule
\begin{equation}
    \mathbf{h}_i^{(l)} = \sigma\Big( \sum_{v_j \in \mathcal{N}_i} \frac{1}{c_{ij}} \mathbf{h}_j^{(l-1)} \mathbf{W}^{(l-1)} \Big)\,,
\end{equation}
where $\mathbf{h}_i^{(l)}$ is the hidden representation of node $v_i$ at layer $l$. Parameters in the weight matrix $\mathbf{W}^{(l-1)}$ are learnable parameters. The quantity $c_{ij}$ is a normalization factor for the edge between $v_i$ and $v_j$. We compute $c_{ij}$ during the data-preprocessing step after we add a self-edge to each node. It is given by
\begin{displaymath}
    c_{ij} = \sqrt{|\mathcal{N}_i| \cdot |\mathcal{N}_j|}\,,
\end{displaymath}
where $\mathcal{N}_i$ is the neighborhood of node $v_i$'s.
The quantity $|\mathcal{N}_i|$ is the size of node $v_i$'s neighborhood.

In the GCN encoder, time information is embedded into the node features. 
Suppose that we have an observed sequence $\{o_i | t < 0\}$ of node $v_i$ with sequence length $\mathrm{SeqEnc}_i$, we compute a weighted sum of all the observations in the sequence.
The weight of the earliest observation is $\frac{1}{2^{\mathrm{SeqEnc}_i}}$, the next observation has weight $\frac{1}{2^{\mathrm{SeqEnc}_i - 1}}$, and so on. The latest observation right before the starting time $t=0$ has weight $1/2$.
The weighted sum of all $\{\mathbf{o}_i^t | t < 0\}$ is the input $\mathbf{h}_i^{(0)}$ to the GCN model.

The last layer of the GCN is followed by a linear-transformation layer with the same architecture as $f_\mathrm{posterior}$ (see Equation~(\ref{eq:encoder_posterior})). 
We then compute the mean $\boldsymbol\mu_i$ and the standard deviation $\boldsymbol\sigma_i$ of the GCN encoder the same way as in Equation~(\ref{eq:encoder_posterior}).

\subsection{The Graph ODE}\label{subsec:ode_solver}

A social-network data set with $N$ accounts, together with update rules of their opinions and the initial states of their opinions, can be viewed as a continuous multi-agent dynamical system.
The opinion evolution of $\mathbf{z}_i^t$ is governed by a set of coupled first-order ODEs with vector fields $g_i$ that update their values in infinitesimal time steps. 

We model our ODE function in a straightforward way using a GNN framework $g_i$ by writing
\begin{equation}
    \frac{\mathrm{d} \mathbf{z}_i^t}{\mathrm{d} t} = g_i(\mathbf{z}_1^{t}, \mathbf{z}_2^{t}, \dots, \mathbf{z}_N^{t}) 
    \,,
\end{equation}
where $g_i$ consists of $L_\mathrm{ODE}$ layers of message-passing convolutional layers:
\begin{equation}
    \mathbf{v}_j^{(l+1)} = f_\mathrm{aggregate}^{(l)} \big( \{ f_\mathrm{message}^{(l)} ( \mathbf{v}_i^{(l)}, \mathbf{v}_j^{(l)} ) | v_i \in \mathcal{N}_j \}\big)\,.
\end{equation}
The input is the nodes' representations $\mathbf{v}_i^{(0)} = \mathbf{z}_i^{t}$ and the output is the derivative $\frac{\mathrm{d} \mathbf{z}_i^t}{\mathrm{d} t} = \mathbf{v}_i^{(L_\mathrm{ODE})}$.

With the initial state $\mathbf{z}_i^0 \in \mathbb{R}^d$ for each account $v_i$, the solution $\mathbf{z}_i^t$ satisfies an initial-value problem (IVP), which we express in the integral form
\begin{equation}
    \mathbf{z}_i^t = \mathbf{z}_i^0 + \int_{t' = 0}^t g_i(\mathbf{z}_1^{t'}, \mathbf{z}_2^{t'}, \dots, \mathbf{z}_N^{t'}\,) \mathrm{d} t' \,.
\end{equation}
There are many numerical ODE solvers to obtain $\mathbf{z}_i^t$ from given initial state $\mathbf{z}_i^0$, time $t$, and ODE function $g_i$. A prominent example is \textsc{Runge--Kutta} methods~\cite{schober2019probabilistic}.
The Euler's method we refer to in our code is a particular type of first-order Runge--Kutta methods~\cite{griffiths2010euler}.

The vector field $g_i$ defines the dynamics of the latent state of the $i$th agent (i.e., account $v_i$). The coupled dynamics of the agents constitute a multi-agent dynamical system.

We have a GNN-based ODE function in our \textbf{\socialdynamical} framework by default (see Section~\ref{subsubsec:graph_ode}), and we also adapt some ODE functions from opinion models to produce other variants (see Section~\ref{subsubsec:opinion_model_ode}).

\subsubsection{Graph Neural-ODE Function}\label{subsubsec:graph_ode}

Our model's ODE function uses a similar updating algorithm as in the \textsc{NRI} model, a type of GNN that has been used successfully to model discrete-time multi-agent dynamical systems~\cite{kipf2018neural}. In each layer of an ordinary \textsc{NRI} model, information is passed from nodes to edges and then from edges to nodes:
\begin{displaymath}
    \mathbf{e}_{(i,j)}^{(l)} = f_e^{(l)}([\mathbf{v}_{i}^{(l)} \mathbin\Vert \mathbf{v}_{j}^{(l)}])\,,\quad \mathbf{v}_{j}^{(l+1)} = f_v^{(l)}(\sum_{i \neq j} \mathbf{e}_{(i,j)}^{(l)})\,,
\end{displaymath}
where $\mathbf{e}_{(i,j)}^{(l)}$ is the edge embedding between node pair $(v_i, v_j)$ in layer $l$ and the vector $\mathbf{v}_{j}^{(l)}$ is the node embedding of node $v_j$ at layer $l$. The functions $f_e^{(l)}$ and $f_v^{(l)}$ are multi-layer perceptrons (MLPs). 
In \textsc{NRI}~\cite{kipf2018neural}, Kpif et al. considered a fully connected graph and discrete time steps. 
We are interested in dynamics on social networks, so we need to consider real-world social-network structure.
Therefore, we adapt the standard \textsc{NRI} node-update rules. For each layer,
\begin{equation}
    \begin{aligned}
        \mathbf{e}_{(i,j)}^{(l)} & = f_\mathrm{message}^{(l)} ( \mathbf{v}_i^{(l)}, \mathbf{v}_j^{(l)} ) = f_e^{(l)}([\mathbf{v}_{i}^{(l)} \mathbin\Vert \mathbf{v}_{j}^{(l)}]) \,,\\
        \mathbf{v}_{j}^{(l+1)} & = f_\mathrm{aggregate}^{(l)}(\{ \mathbf{e}_{(i,j)}^{(l)} | v_i \in \mathcal{N}_j \}) = f_v^{(l)}(\sum_{v_i \in \mathcal{N}_j} [ \mathbf{v}_{i}^{(l)} \mathbin\Vert \mathbf{e}_{(i,j)}^{(l)}] ) + \mathbf{v}_{j}^{(l)} \,.
    \end{aligned}
\end{equation}
Following settings that have worked well in similar models~\cite{kipf2018neural,NEURIPS2020_ba484941}, we set the number of layers in our ODE function $L_\mathrm{ODE}$ to $1$ as a default. 
When we change it to other values, we find the model performance degrades.

\subsubsection{Baseline Opinion Models}\label{subsubsec:opinion_model_ode}

In some other variants of \textbf{\socialdynamical}, we adapt the right-hand sides of our ODE functions from opinion models to GNN models. 
In our discussion, we use the \textsc{SINN}~\cite{okawa2022predicting} baseline models because they are already neural-network models and worked well on some simulated data sets. These baselines include
the \textsc{DeGroot} model~\cite{abelson1964mathematical}, the \textsc{Friedkin--Johnsen} (FJ) model~\cite{friedkin1990social}, and a \textsc{Hegselmann--Krause} (HK) bounded-confidence model~\cite{rainer2002opinion}.

\begin{itemize}
    \item Among all opinion models we have translated into GNN model, the \textsc{DeGroot} model is the simplest one with the fewest constraints and requires the least modification from \textsc{SINN} model to GNN model. 
    Denote latent opinion of an account $v_i$ at time $t$ by $\mathbf{z}_i^t \in \mathbb{R}^d$. Its discrete-time update rule on a network is
    \begin{displaymath}
        \mathbf{z}_j^{t+1} = \mathbf{z}_j^t + \sum_{v_i \in \mathcal{N}_j} a_{ij} \mathbf{z}_i^t \,,
    \end{displaymath}
    where $\mathcal{N}_j$ is the neighborhood of node $v_j$ and $a_{ij} > 0$ is the strength of the edge (i.e., the follower--followee relationship) between two adjacent accounts $v_i$ and $v_j$. The \textsc{SINN} model uses an ODE as a continuous-time analogue of the \textsc{DeGroot} model in the form
    \begin{displaymath}
        \frac{\mathrm{d} \mathbf{z}_i^t}{\mathrm{d} t} = \sum_{v_i \in \mathcal{N}_j} a_{ij} \mathbf{z}_i^t = \sum_{v_i \in \mathcal{N}_j} \mathbf{m}_i^T \mathbf{q}_j\mathbf{z}_i^t\,,
    \end{displaymath}
    where $\mathbf{m}_i, \mathbf{q}_j \in \mathbb{R}^K$ are the $i$th and $j$th columns of $\mathbf{M}, \mathbf{Q} \in \mathbb{R}^{N \times K}$.
    The number of nodes of the network is $N$, and we use $\mathbf{M}$ and $\mathbf{Q}$ to decompose the $N \times N$ amount of parameters of $a_{ij}$ with the expression $\mathbf{a} = \mathbf{M} \mathbf{Q}^T$.
    The \textsc{SINN} model sets the hyperparameter $K \ll N$, which dramatically reduces the number of parameters from $\mathcal{O}(N^2)$ to $\mathcal{O}(KN)$.
    Translating this formula into a GNN model yields
    \begin{equation}
    \begin{aligned}
        \mathbf{e}_{(i,j)}^{(l)} & = f_\mathrm{message}^{(l)} ( \mathbf{v}_i^{(l)}, \mathbf{v}_j^{(l)} ) = \mathbf{m}_i^T \mathbf{q}_j\mathbf{v}_i^{(l)}\,, \\
        \mathbf{v}_{j}^{(l+1)} & = f_\mathrm{aggregate}^{(l)}(\{ \mathbf{e}_{(i,j)}^{(l)} | v_i \in \mathcal{N}_j \}) = \sum_{v_i \in \mathcal{N}_j} \mathbf{e}_{(i,j)}^{(l)}\,.
    \end{aligned}
    \end{equation}
    When $L_\mathrm{ODE} = 1$, our graph-ODE function is identical to the \textsc{SINN} \textsc{DeGroot} ODE function.

    \item The \textsc{Friedkin--Johnsen} (FJ) model is an opinion model that considers accounts with different susceptibilities to interpersonal influence.
    Some researchers also use the term ``stubborness'' to describe FJ model's incorporation of nodes with hesitance to change their opinions~\cite{wang2021achieving}. The discrete-time FJ update rule is
    \begin{displaymath}
        \mathbf{z}_j^{t+1} = (1 - s_j) \mathbf{z}_j^0 + s_j \sum_{v_i \in \mathcal{N}_j} \mathbf{z}_i^t \,,
    \end{displaymath}
    where $s_j \in [0, 1]$ is the susceptibility to persuasion. A smaller value of $s_j$ signifies that account $v_j$ is harder for others to influence. 
    All neighbors in a network have the same importance. Okawa and Iwata~\cite{okawa2022predicting} proposed an ODE that is similar to the FJ model. It is
    \begin{displaymath}
        \frac{\mathrm{d} \mathbf{z}_i^t}{\mathrm{d} t} = s_j \sum_{v_i \in \mathcal{N}_j} \mathbf{z}_i^t + (1 - s_j) \mathbf{z}_j^0 - \mathbf{z}_j^t \,.
    \end{displaymath}
    When $L_\mathrm{ODE} = 1$, by extending the dimensionality of $\mathbf{z}_j^t$ from one dimension ($1$D) to $d$ dimensions ($d$D), we obtain a GNN version of \textsc{FJ} ODE. We do not include the decay term $- \mathbf{z}_j^t$. This yields our GNN FJ model
    \begin{equation}
    \begin{aligned}
        \mathbf{e}_{(i,j)}^{(l)} & = f_\mathrm{message}^{(l)} ( \mathbf{v}_i^{(l)}, \mathbf{v}_j^{(l)} ) = \mathbf{s}_j \odot \mathbf{v}_i^{(l)} \,, \\
        \mathbf{v}_{j}^{(l+1)} & = f_\mathrm{aggregate}^{(l)}(\{ \mathbf{e}_{(i,j)}^{(l)} | v_i \in \mathcal{N}_j \}) = \sum_{v_i \in \mathcal{N}_j} \mathbf{e}_{(i,j)}^{(l)} + (\mathbbm{1} - \mathbf{s}_j) \odot \mathbf{v}_j^{(0)} \,,
    \end{aligned}
    \end{equation}
    where $\mathbf{x} \odot \mathbf{y}$ denotes element-wise multiplication. For any node $v_j$, the susceptibility value and the hidden value of the node (in each layer of the GNN function) have the same dimensionality (i.e., $\mathbf{s}_j, \mathbf{v}_j^{(l)} \in \mathbb{R}^d$). 
    
    \item The \textsc{Hegselmann--Krause} (HK) model is a bounded-confidence model (BCM)~\cite{bernardo2024bounded}. Bounded-confidence models consider {``confirmation bias''} and ``selective exposure'', which entails that agents tend to pay more attention to information that confirms their preconceptions~\cite{schmidt2017anatomy,chitra2020analyzing}. The discrete-time HK model is
    \begin{displaymath}
        \mathbf{z}_j^{t+1} = \mathbf{z}_j^t + \frac{1}{|{N}_j(t)|} \sum_{i \in {\Gamma}_j(t)} (\mathbf{z}_i^t - \mathbf{z}_j^t) \,,
    \end{displaymath}
    where ${\Gamma}_j(t) = \{ v_i \in \mathcal{N}_j \ | \ |\mathbf{z}_i^t - \mathbf{z}_j^t| \leq \delta \}$ is the set of neighboring nodes whose opinions are within the confidence bound of the target node $j$ (i.e., $|\mathbf{z}_i^t - \mathbf{z}_j^t| \leq \delta$). 
    Okawa and Iwata~\cite{okawa2022predicting} used a continuous-time HK model in the form of the ODE
    \begin{displaymath}
        \frac{\mathrm{d} \mathbf{z}_i^t}{\mathrm{d} t} = \sum_{v_i \in \mathcal{N}_j} \sigma\big( \delta -  |\mathbf{z}_i^t - \mathbf{z}_j^t| \big) \big( \mathbf{z}_i^t - \mathbf{z}_j^t \big)\,,
    \end{displaymath}
    where $\mathbf{z}_j^t \in \mathbb{R}^1$ and $\sigma(z) = 1 / ( 1 + e^{\gamma z})$. 
    In our scenario, we extend the agent opinions from $1$D to $d$D, with $\mathbf{z}_j^t \in \mathbb{R}^d$, and assume that an agent's attention to all different topics (with one topic in each dimension of the $d$-dimensional opinion space) is limited. Therefore, we use the softmax smooth function and write
    \begin{equation}
    \begin{aligned}
        \mathbf{e}_{(i,j)}^{(l)} & = f_\mathrm{message}^{(l)} ( \mathbf{v}_i^{(l)}, \mathbf{v}_j^{(l)} ) = \boldsymbol\gamma \odot \mathrm{Softmax}\big( \boldsymbol\xi \odot (\boldsymbol\delta - |\mathbf{v}_i^t - \mathbf{v}_j^t|) \big) \odot \big( \mathbf{v}_i^t - \mathbf{v}_j^t \big)\,, \\
        \mathbf{v}_{j}^{(l+1)} & = f_\mathrm{aggregate}^{(l)}(\{ \mathbf{e}_{(i,j)}^{(l)} | v_i \in \mathcal{N}_j \}) = \sum_{v_i \in \mathcal{N}_j} \mathbf{e}_{(i,j)}^{(l)}\,,
    \end{aligned}
    \end{equation}
    where $\odot$ represents element-wise multiplication and $\boldsymbol\xi, \boldsymbol\delta, \boldsymbol\gamma \in \mathbb{R}^d$, which are the same for all agents, are parameters to learn. 
    In practice, on our real-world data sets, we found softmax smoothing function performs significantly better than the sigmoid function that \textsc{SINN} used for $1$D opinion updates.
\end{itemize}

To demonstrate the importance of the ODE functions, we also use a naive baseline ODE model, which we call \textsc{No-Update}. This model is
\begin{equation}
    \begin{aligned}
        \mathbf{e}_{(i,j)}^{(l)} & = f_\mathrm{message}^{(l)} ( \mathbf{v}_i^{(l)}, \mathbf{v}_j^{(l)} ) = \mathbf{v}_j^{(l)} \,,\\
        \mathbf{v}_{j}^{(l+1)} & = f_\mathrm{aggregate}^{(l)}(\{ \mathbf{e}_{(i,j)}^{(l)} | v_i \in \mathcal{N}_j \}) = \frac{1}{|\mathcal{N}_j|}\sum_{v_i \in \mathcal{N}_j} \mathbf{e}_{(i,j)}^{(l)}\,.
    \end{aligned}
\end{equation}
In this case, the ODE has the same opinion prediction as the initial opinion $\mathbf{v}_j^0$ for each node $v_j$ for all times $t$. 

\subsection{The Time-Dependent Decoder}\label{sec:dynamic_decoder}

We consider two tasks: (1) interaction inference (i.e., ``prediction'') and (2) polarity inference (i.e., ``prediction''). 
As we discussed in Section~\ref{sec:opinion_eval}, each of the downstream tasks has a different decoder design and different objectives.

Previous neural dynamical systems~\cite{kipf2018neural,NEURIPS2020_ba484941} jointly train the encoder, a reconstruction decoder, and an ODE by maximizing the evidence lower bound (ELBO)
\begin{equation}\label{eq::elbo}
    \mathrm{ELBO}(\theta, \phi) = \mathbb{E}_{\mathbf{Z^0}\sim q_\phi(\mathbf{Z^0} | o_1, o_2, \dots, o_N)} [\log p_\theta(o_1, o_2, \dots, o_N)] - \mathrm{KL} [q_\phi (\mathbf{Z^0} | o_1, o_2, \dots, o_N) \mathbin\Vert p(\mathbf{Z^0})]\,,
\end{equation}
where the prior $p(\mathbf{Z^0})$ is the standard normal distribution $\mathcal{N}(\mymathbb{0}, \mymathbb{1})$. 
The vectors $\mymathbb{0}$ and $\mymathbb{1}$ are all-zero and all-one vectors, respectively.
The quantity $\mathrm{KL}[q_\phi \mathbin\Vert p]$ is the Kullback--Leibler (KL) divergence between the distribution $q_\phi$ and the distribution $p$~\cite{zhang2024properties}.
Previous neural dynamical systems have considered an observation-reconstruction task~\cite{kipf2018neural,NEURIPS2020_ba484941}. However, real-world social networks involve many uncertainty. It is unreasonable to suppose that we can predict exactly what the accounts post. Therefore, we seek different decoder tasks than observation--reconstruction.

We use a similar framework as in Equation~(\ref{eq::elbo}). We break the loss value into two parts. The first part estimates the quality of the decoder's prediction, and the second part is a regularization term. The loss value is
\begin{equation}\label{eq::objective}
    \mathcal{L} = \mathcal{L}_\mathrm{dec} + \lambda \mathcal{L}_\mathrm{reg} \,,
\end{equation}
where $\mathcal{L}_\mathrm{reg} = \mathrm{KL} [q_\phi (\mathbf{Z^0} | o_1, o_2, \dots, o_N) \mathbin\Vert p(\mathbf{Z^0})]$. The term $\mathcal{L}_\mathrm{dec}$ is different for different tasks.

\subsubsection{Interaction Prediction}

We model the interaction-prediction task as a time-dependent edge-prediction problem.
We assign an index $R \in \{ 1, 2, 3, 4 \}$ to the $4$ types of possible interactions: {reply}, {mention}, {retweet}, and {like}.
For a temporal edge $\langle i, j, t\rangle$ with relation $r \in \{ 1, 2, 3, 4 \}$, node indices $i, j \in \{1, 2, \dots, N\}$, and time $t \in [0, T]$ (where $T$ is the ending time of the observations), we measure the score of the edge's existence with the function
\begin{equation}\label{eq::score_temp_edge}
    \mathrm{edge\_score}(i, j, t, r) = \mathbf{z}_i^t \mathbf{W}_r \mathbf{z}_j^t + \mathbf{W}_v \begin{bmatrix} \mathbf{z}_i^t \\ \mathbf{z}_j^t \end{bmatrix} + b\,.
\end{equation}
This choice extends the \textsc{TIMME-NTN} module~\cite{10.1145/3394486.3403275} for edge prediction to temporal edges. In Equation~(\ref{eq::score_temp_edge}), $\mathbf{W}_r \in \mathbb{R}^{d \times d}$, $\mathbf{W}_v \in \mathbb{R}^{2d}$, and $b \in \mathbb{R}$ are trainable parameters of the decoder. The matrix $\mathbf{W}_r$ is diagonal.

When the temporal edge $\langle i, j, t\rangle$ exists, the ground-truth value of the edge score is $1$; when it does not exist, the ground-truth value is $0$.
We optimize $\mathrm{edge\_score}(i, j, t, r)$ using the ground-truth values by minimizing the binary-cross-entropy loss, which we thus use for $\mathcal{L}_\mathrm{dec}$. We evaluate the models' performance on the test data set by using the ROC-AUC and by the mean precision. 
The mean precision is the area under the curve (AUC) of the precision--recall (PR) curve, so we also refer to it as PR-AUC. The ROC-AUC score means the Receiver Operating Characteristic Area Under the Curve.

\subsubsection{Polarity Prediction}

We model the polarity-prediction task as a regression problem. We label the ground-truth political-polarity score using the polarity scores that we compute with \textsc{PEM}~\cite{xiao2023detecting}. 

We feed $x_i^t$ into the pipeline of the pretrained \textsc{PEM} model sentence by sentence. We calculate the mean value of the polarity score of all tokens in a sentence as a sentence-level polarity score. We regard the mean value of all sentences' polarity scores as a ground-truth polarity score $p_i^t$.

For any text-content observation $x_i^t$ of account $v_i$ at time $t > 0$, the corresponding latent opinion observation $\mathbf{z}_i^t$ is available. 
We then use a $2$-layer MLP network and output a $1$D score
\begin{displaymath}
     \hat{p}_i^t = \mathrm{MLP}(\mathbf{z}_i^t) 
\end{displaymath}
as the predicted polarity value.
We compute the loss component $\mathcal{L}_\mathrm{dec}$ as a negative Gaussian log-likelihood between the prediction $\hat{p}_i^t \in \mathbb{R}$ and the ground truth $p_i^t \in \mathbb{R}$.
We evaluate the quality of our prediction by calculating the mean-square error (MSE) and mean absolute-percentage error (MAPE) between all pairs of $\hat{p}_i^t$ and $p_i^t$.

\subsubsection{Polarity Classification}

We also tried to model the polarity prediction task as a classification problem, labeled by the political-polarity scores computed by \textsc{PEM}~\cite{xiao2023detecting}. 
For any text content observation $x_i^t$ of account account $i$ at time $t > 0$, the corresponding latent opinion observation $\mathbf{z}_i^t$ is available. Then we use a $2$-layer MLP network and output a $2$-D score, followed by a softmax layer that helps to select the most likely class:
\begin{displaymath}
    \hat{p}_i^t = \arg\max_{j \in \{ 0, 1 \} } [\mathrm{Softmax}_j \big(\mathrm{MLP}(\mathbf{z}_i^t) \big) ]\,,
\end{displaymath}
where $\mathrm{Softmax}_j$ represents the $j^{th}$ element of $\mathrm{Softmax}\big(\mathrm{MLP}(\mathbf{z}_i^t) \big) \in \mathbb{R}^2$, meaning that when the first (i.e. index $0$) unit in the output is higher, class label $0$ is chosen, otherwise $1$. We feed $x_i^t$ into the pipeline of \textsc{PEM} model sentence after sentence and take a mean of the polarity scores of every sentence in $x_i^t$. The mean value of all sentences' polarity scores is regarded as the raw ground-truth polarity score $\tilde{p}_i^t$. We then define:
\begin{displaymath}
    p_i^t =
    \begin{cases}
        1 & \tilde{p}_i^t > 0 \\
        0 & \text{otherwise}
    \end{cases}
\end{displaymath}
The case where $\tilde{p}_i^t$ is neglectable because, according to the pre-trained \textsc{PEM} model, all non-empty sentences could obtain a non-zero political polarity.

Then we compute the $\mathcal{L}_\mathrm{dec}$ as a negative log likelihood between prediction $\hat{p}_i^t \in \{ 0, 1 \}$ and ground truth $p_i^t \in \{ 0, 1 \}$.
We evaluate the quality of the prediction by measuring the accuracy and F1-scores between all pairs of $\hat{p}_i^t$ and $p_i^t$.

\subsubsection{Text-Embedding Reconstruction}

The text-embedding reconstruction task is an unsupervised task where no label is needed. We have the ground-truth observation $\mathbf{o}_i^t$ of any node $v_i$ at any time $t$, and the corresponding latent opinion representation $\mathbf{z}_i^t$ . By passing $\mathbf{z}_i^t$ through a single linear layer that does a linear transformation, and adapts the output dimension to the size of $\mathbf{o}_i^t$, we have a predicted observation $\hat{\mathbf{o}}_i^t$:
\begin{displaymath}
    \hat{\mathbf{o}}_i^t = \mathrm{Linear}(\mathbf{z}_i^t) \,,
\end{displaymath}
and the negative Gaussian log likelihood between prediction $\hat{\mathbf{o}}_i^t$ and ground truth $\mathbf{o}_i^t$ is used as the objective $\mathcal{L}_\mathrm{dec}$. 
We evaluate the quality of the reconstruction process by taking the mean value of all the MSE (Mean Square Error) scores between any pair of $\hat{\mathbf{o}}_i^t$ and $\mathbf{o}_i^t$.

%% file: 5-experiments.tex
\section{Experiments}\label{sec:sds_experiments}

In this section, we show how we designed our experiments in details, and discuss the experimental results of our \textbf{\socialdynamical} framework.

\subsection{Data Sets}\label{subsec::sds_dataset}

We collected our data set from Twitter using the Twitter (currently named $\mathbb{X}$) API.\footnote{See \url{https://developer.twitter.com/en/docs/twitter-api} for more information.} We collected the data by the end of the year 2020.\footnote{The last valid record of post was on November $26$th.} Back then, the Twitter API allowed access to the most recent $3{,}200$ tweets in each account's timeline. 

Although some accounts' records can be traced back to 2019 or even earlier, we align the starting time and the ending time of their timelines and keep only the tweets that were posted in 2020.

\subsubsection{Constructing Dynamic Graphs}\label{data_prep:accounts_selection}

We use the \textsc{TIMME} data set collection of politicians' Twitter accounts~\cite{10.1145/3394486.3403275}, which includes members of Congress, cabinet officials, and presidential candidates at that time. After filtering out unavailable accounts and those with fewer than two tweets in 2020, we retain $513$ politician accounts.

In our dynamic network, each node represents a Twitter account, while edges denote follower–followee relationships, which we assume remain static throughout the year, while opinions evolve over time.

To train our model effectively, we construct $50$ dynamic graph sequences from the 2020 tweets, following best practices from previous neural-ODE studies~\cite{NEURIPS2020_ba484941}. Each sequence consists of all the $513$ politicians' accounts and $287$ randomly-selected non-politicians from the TIMME data set. The node sets vary between sequences, except for the politicians, who are included in every sequence to maintain graph structure connectivity.

Each sequence spans an entire year. Although our model operates on continuous time and can, in theory, capture minute- or second-level opinion shifts, we assume human opinions change more gradually. Thus, we set a temporal granularity of one week (seven days).

\subsubsection{From Observations to Representations}\label{sec:obs_to_sentence_bert}

What we observed at time $t$ for account $i$ is not directly possessed in vector space. Instead of having a ground-truth representation $\mathbf{o}_i^t$, we observe the text $x_i^t$ that an account $v_i$ posts on Twitter at time $t$.
For our tasks, we use the \textsc{Sentence-BERT} model~\cite{reimers2019sentence}. We use the {``all-MiniLM-L6-v2''} version of its pretrained parameters to translate text sentences into vector embeddings. 
The \textsc{Sentence-BERT} model has been tested on benchmarks such as \textsc{SentEval}~\cite{conneau2018senteval}, and it has performed better on sentence-level embeddings~\cite{reimers2019sentence} than other popular language models such as \textsc{BERT}~\cite{devlin2018bert} and \textsc{RoBERTa}~\cite{liu2019roberta}.

We transform the observations $\tilde{X}_i^t = \{x_i^{\tilde{t}} | \tilde{t} \in [t_\mathrm{start}, t_\mathrm{end}] \}$ of account $v_i$'s tweets within a certain week into vector-space representations 
$\tilde{O}_i^t = \{\tilde{\mathbf{o}}_i^{\tilde{t}} | \tilde{t} \in [t_\mathrm{start}, t_\mathrm{end}] \}$, 
and we then take a mean of these representations. This yields our vector-space representation of account $v_i$ at week $t$. It is given by
\begin{equation}
    \mathbf{o}_i^t = \frac{1}{|\tilde{O}_i^t|} \sum_{ \tilde{t} \in [t_\mathrm{start}, t_\mathrm{end}] } \tilde{\mathbf{o}}_i^{\tilde{t}}\,.
\end{equation}

\subsubsection{Limitations}\label{subsubsec::dataset_limitations}

\begin{figure}[!htbp]
    \centering
    \includegraphics[width=0.8\linewidth]{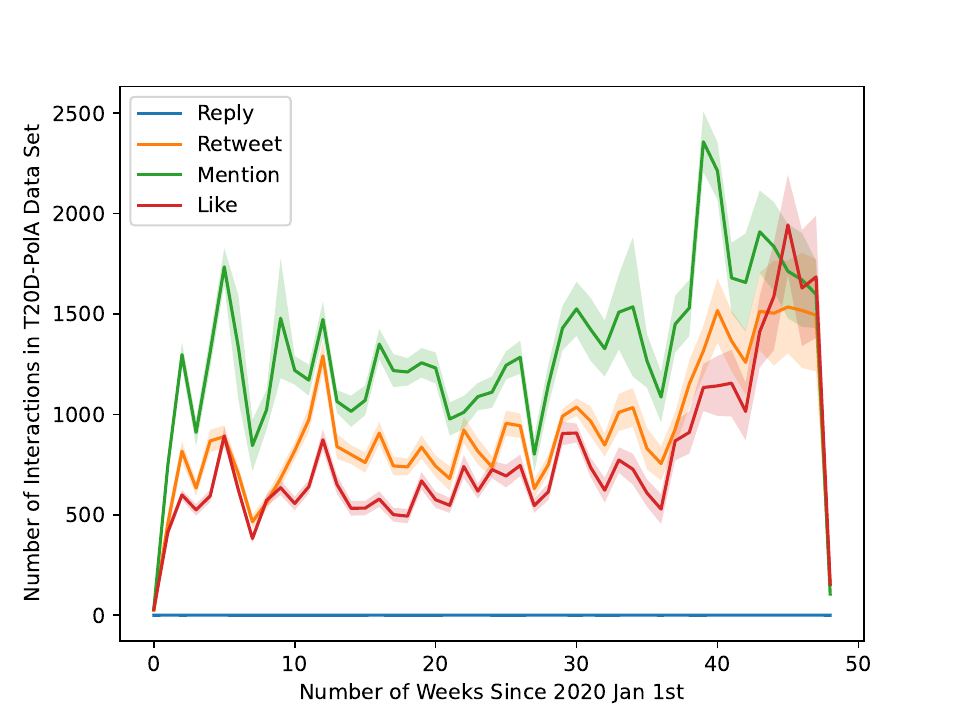}
    \caption{The number of interactions observed in our data set throughout year 2020, from the first week to the last. The average number of total replies per day in each week consistently remains below $1$. The solid line represents the mean values, while the shaded areas indicate the standard deviation.}
    \label{fig:dataset_interactions}
\end{figure}

\begin{figure}[!htbp]
    \centering
    \includegraphics[width=0.8\linewidth]{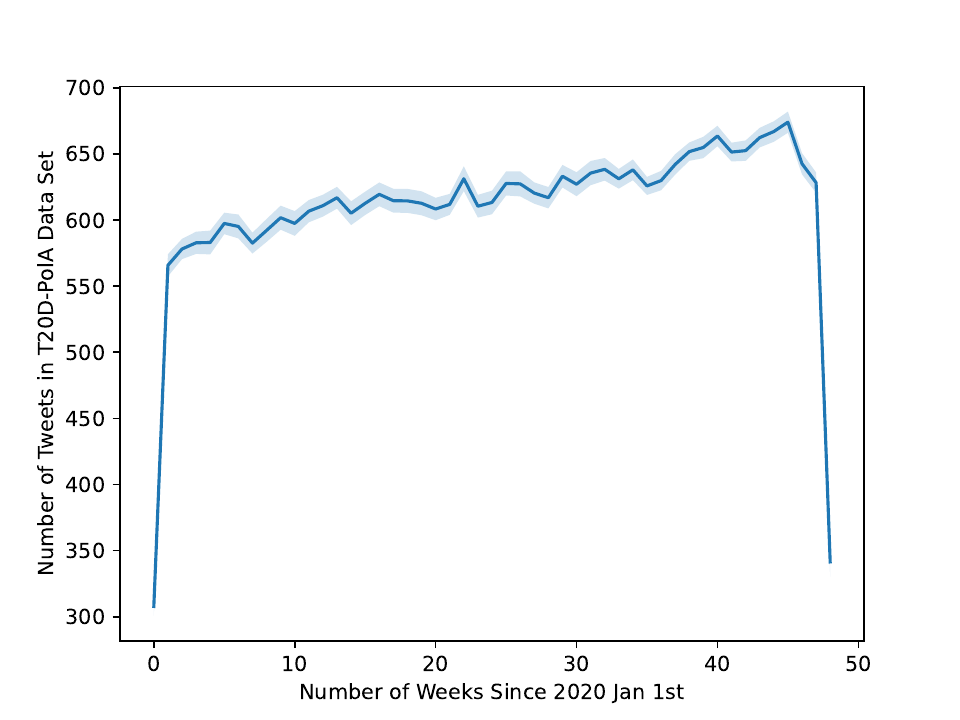}
    \caption{The number of tweets included in our data set throughout year 2020, from the first week to the last. There are significantly fewer observations in the first and last weeks because these weeks are incomplete, meaning not all days are fully observed. The solid line represents the mean values, while the shaded areas indicate the standard deviation.}
    \label{fig:dataset_observed_tweets}
\end{figure}

We examine the numbers of observations each week for replies, mentions, retweets, and likes. We plot their mean values and the standard deviations in Figure~\ref{fig:dataset_interactions}. We also plot the number of tweets observed each week, plotting the mean values and standard deviations in Figure~\ref{fig:dataset_observed_tweets}.

Due to the accessibility of the Twitter API, more recent behaviors are preserved more effectively compared to earlier ones. This phenomenon is particularly evident in the number of interactions (e.g., retweets) between user accounts (see Figure~\ref{fig:dataset_observed_tweets}), as many older tweets are no longer accessible.

\begin{figure}[!htbp]
    \centering
    \includegraphics[width=0.8\linewidth]{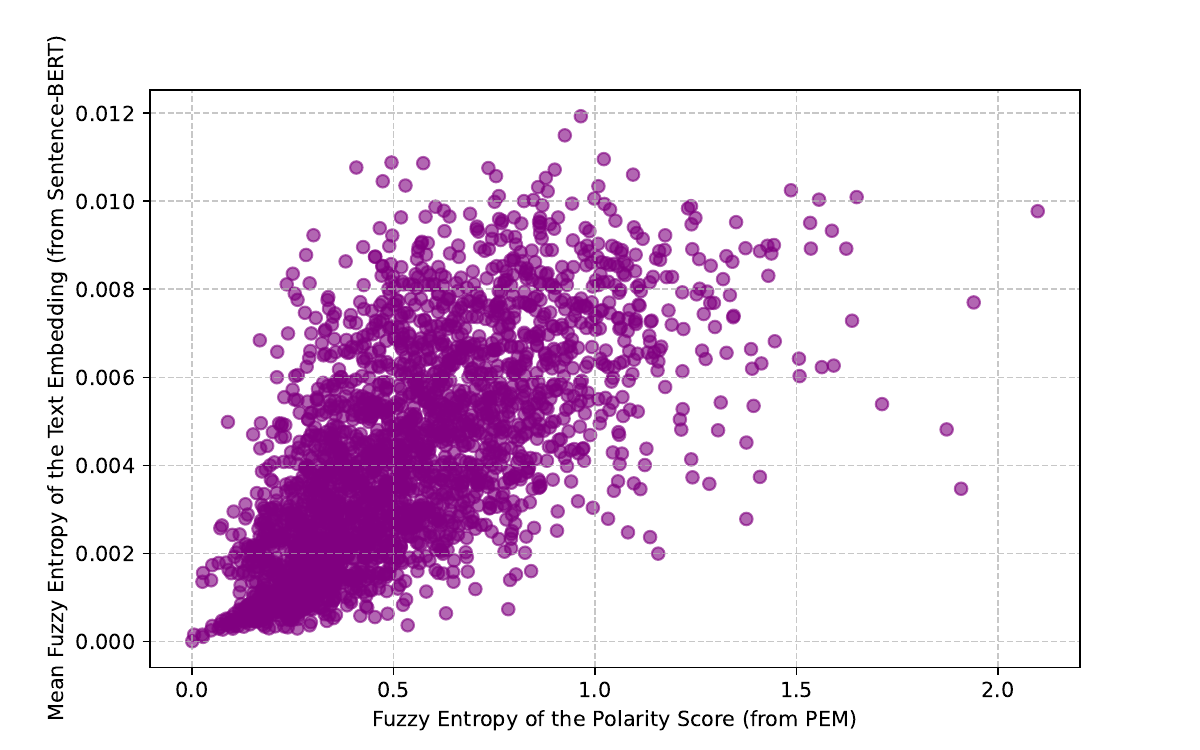}
    \caption{The mean Fuzzy Entropy scores of each account's text embedding values of all dimensions from \textsc{Sentence-BERT}~\cite{reimers2019sentence} embedding, and the mean Fuzzy Entropy scores of the same account's polarity scores from \textsc{PEM}~\cite{xiao2023detecting}.}
    \label{fig:dataset_fluctuation}
\end{figure}

Another limitation lies in the lack of fluctuations in the text features extracted by \textsc{Sentence-BERT}~\cite{reimers2019sentence} (see Section~\ref{sec:obs_to_sentence_bert}). 
To show this limitation, we use Fuzzy entropy to evaluate our data set~\cite{borin2021multiscale}.
Fuzzy entropy is a measure of uncertainty within the framework of fuzzy set theory. It quantifies the degree of fuzziness in data, helping to analyze ambiguity or vagueness in a given data set or model. Fuzzy entropy can also be used to assess the complexity or irregularity of data, including the magnitude of fluctuations. It is widely applied in areas such as signal processing, biomedical analysis, and time-series analysis.
As we can see from Figure~\ref{fig:dataset_fluctuation}, compared to the Fuzzy entropy of the polarity scores measured by \textsc{PEM}~\cite{xiao2023detecting} embeddings' polarity components, the Fuzzy entropy of the features we extracted from \textsc{Sentence-BERT} is significantly lower. This phenomenon indicates that the fluctuation of political polarity tendencies are not significantly reflected in the change of the text embeddings. However, from Figure~\ref{fig:dataset_fluctuation}, we also see that the fuzzy entropies of an account's political polarity score and its text embeddings are somewhat positively correlated in general. It suggests that \textsc{Sentence-BERT} embeddings do have the potential of capturing the change of opinions.

Overall, as shown in Figure~\ref{fig:dataset_fluctuation}, we observe that the fuzzy entropy values are generally low, indicating that detecting subtle changes in opinions over time is a challenging task. This is not surprising, as previous research has long established that people tend to interpret new events or reinterpret past events in ways that maintain consistency with their existing predispositions~\cite{abelson1959modes,lenz2009learning}.
However, the limited fluctuation in opinions we observed adds to the difficulty of evaluating the effectiveness of our ODE functions, as both the best and worst-performing models may achieve similar outcomes in learning the updating rules.

\subsection{Predicting Future Outcome}\label{sec:performance}

\input{table-performance}

In Table~\ref{tab:performance}, we present the performance of our \textbf{\socialdynamical} model on the downstream tasks discussed in Section~\ref{sec:opinion_eval}, using the decoder architectures described in Section~\ref{sec:dynamic_decoder}.
By default, the \textbf{\socialdynamical} model refers to the architecture that incorporates the temporal-graph encoder from Section~\ref{sec:encoder} and the NRI-based graph ODE function from Section~\ref{subsubsec:graph_ode}.

Additionally, we report the performance of model variants that use different encoders or ODE functions. Comparing different variants of the \textbf{\socialdynamical} model serves as an ablation study, helping to assess the contributions of each component.

We observe that the choice of encoder has a significantly greater impact on performance compared to the choice of ODE functions in our data set.
This is likely related to the limitations of our data set collected from Twitter, as discussed in Section~\ref{subsubsec::dataset_limitations}. The relatively high performance of the \textsc{No-Update} setting, which assumes that opinions remain unchanged, suggests that the behaviors of the selected accounts exhibit consistency over time within the observed time span.

Polarity classification task is the least challenging task. Although text-embedding reconstruction task is also very easy, evidence is that all variants of the model achieve very high performance, the terrible performance of \textsc{No-Hidden} variant of our model, which do not learn an initial hidden representation $\mathbf{z}_i^0$ for node $i$, instead, it uses the text embedding directly --- its terrible performance indicates that the text-embedding task does require a good encoder to achieve reasonable performance.

The polarity classification task is the least challenging among all. Although the text-embedding reconstruction task is also relatively easy—evidenced by the high performance of all model variants—the poor performance of the \textsc{No-Hidden} variant suggests otherwise. This variant, which does not learn an initial hidden opinion representation $\mathbf{z}_i^0$ for node $i$ and instead directly uses $\mathbf{o}_i^0$, the last text embedding before the starting time, performs significantly worse. This indicates that, despite its apparent simplicity, the text-embedding reconstruction task still requires a well-designed encoder to achieve reasonable performance.

Among the four decoder tasks we explored, the interaction-prediction task is the most challenging. The performance differences among various model variants are most pronounced in this task. On the interaction-prediction task, our \textbf{\socialdynamical} model achieves the best performance.

\subsection{Predicting Longer-Term Outcomes}\label{sds:case_study}

While the performance reported in Section~\ref{sec:performance} provides a general summary that treats all time steps equally, we are also interested in evaluating how well our model can predict longer-term future outcomes.

\begin{figure}[!htbp]
    \centering
    \includegraphics[width=0.95\linewidth]{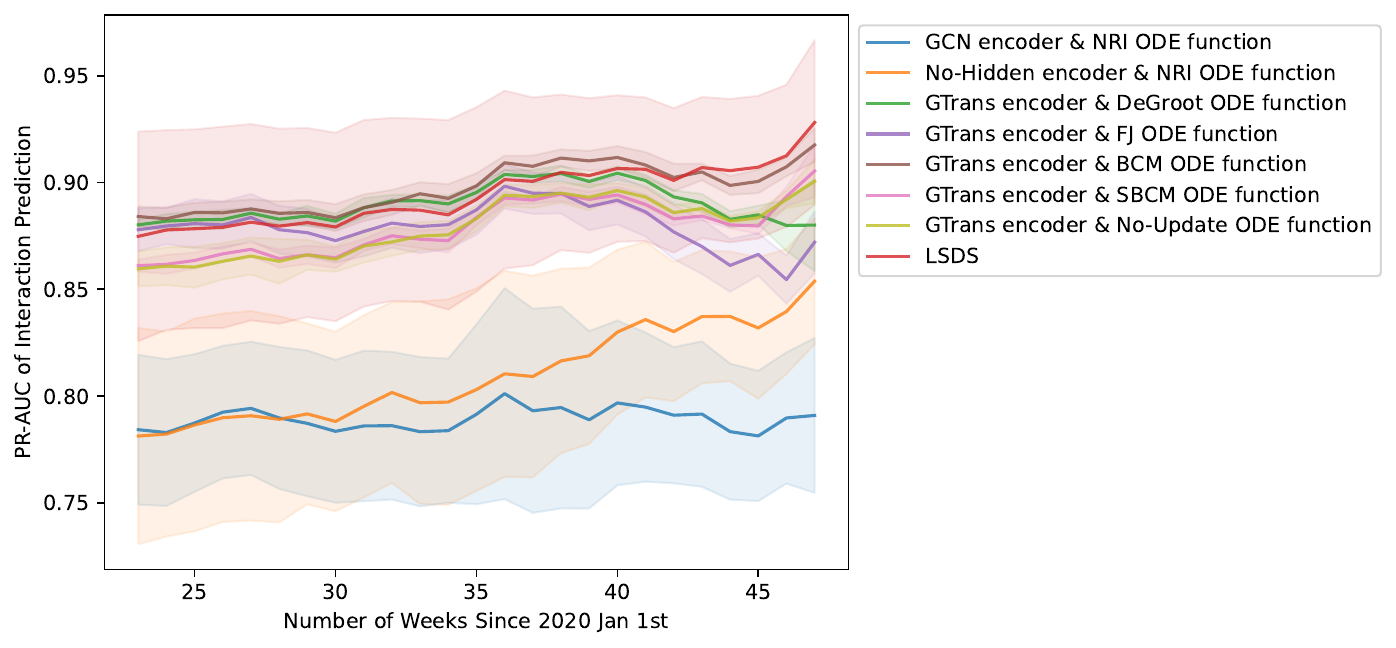}
    \caption{Performance of the interaction-prediction task for different prediction lengths. The encoder uses the first half of the observation sequence, while the second half is used for prediction. The solid line represents the mean values, while the shaded areas indicate the standard deviation.}
    \label{fig:future_interaction}
\end{figure}

\begin{figure}[!htbp]
    \centering
    \includegraphics[width=0.95\linewidth]{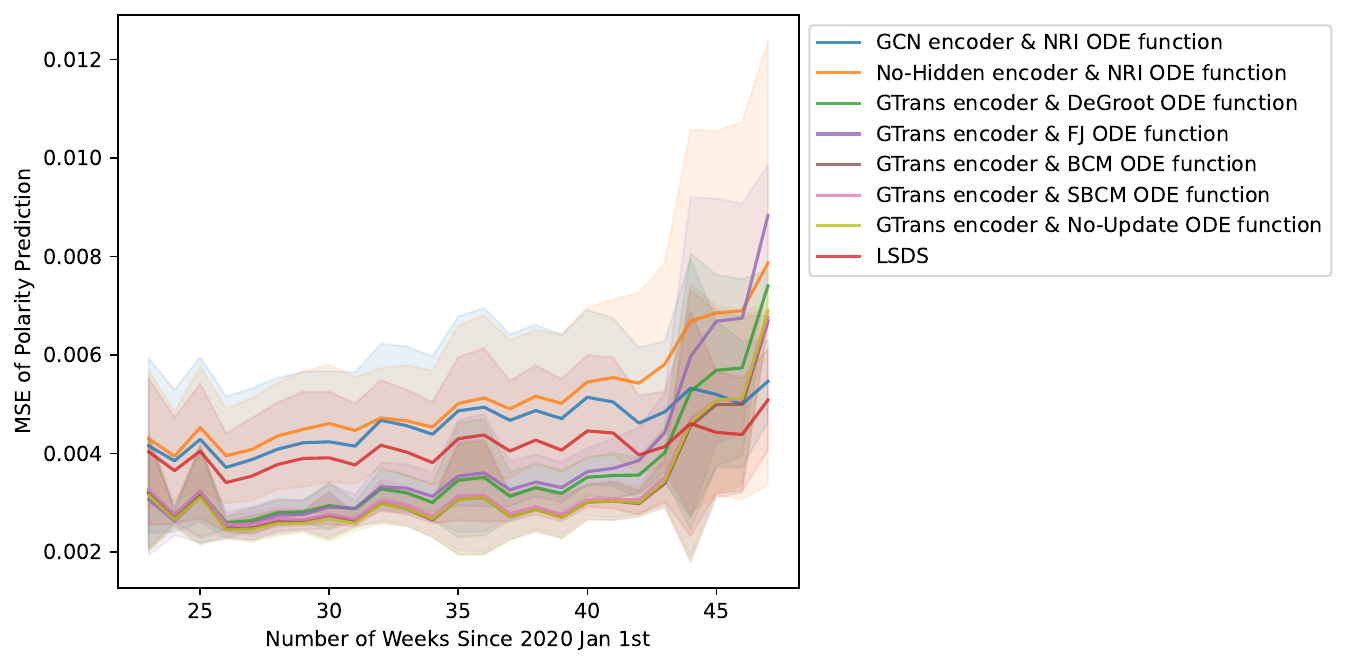}
    \caption{Errors of the polarity-prediction task for different prediction lengths. The encoder uses the first half of the observation sequence, while the second half is used for prediction. The solid line represents the mean values, while the shaded areas indicate the standard deviation.}
    \label{fig:future_polarity_pred}
\end{figure}

\begin{figure}[!htbp]
    \centering
    \includegraphics[width=0.95\linewidth]{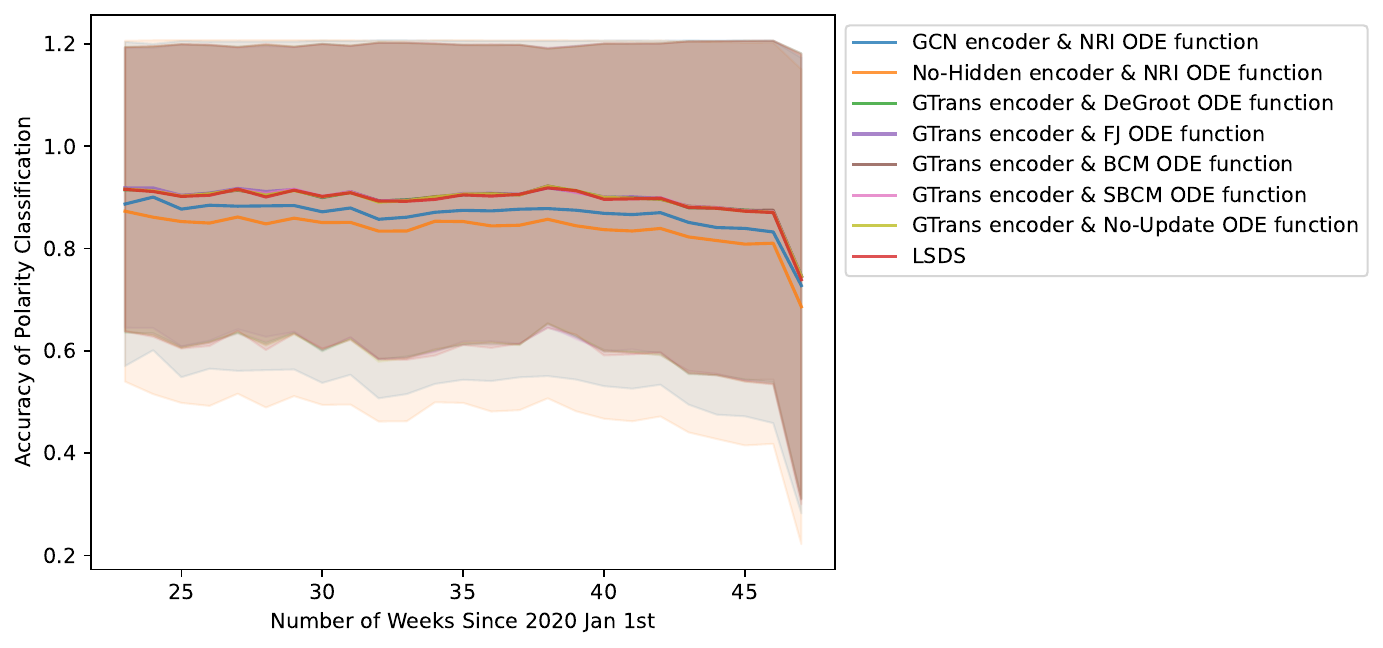}
    \caption{Performance of the polarity-classification task for different prediction lengths. The encoder uses the first half of the observation sequence, while the second half is used for prediction. The solid line represents the mean values, while the shaded areas indicate the standard deviation.}
    \label{fig:future_polarity_classification}
\end{figure}

\begin{figure}[!htbp]
    \centering
    \includegraphics[width=0.95\linewidth]{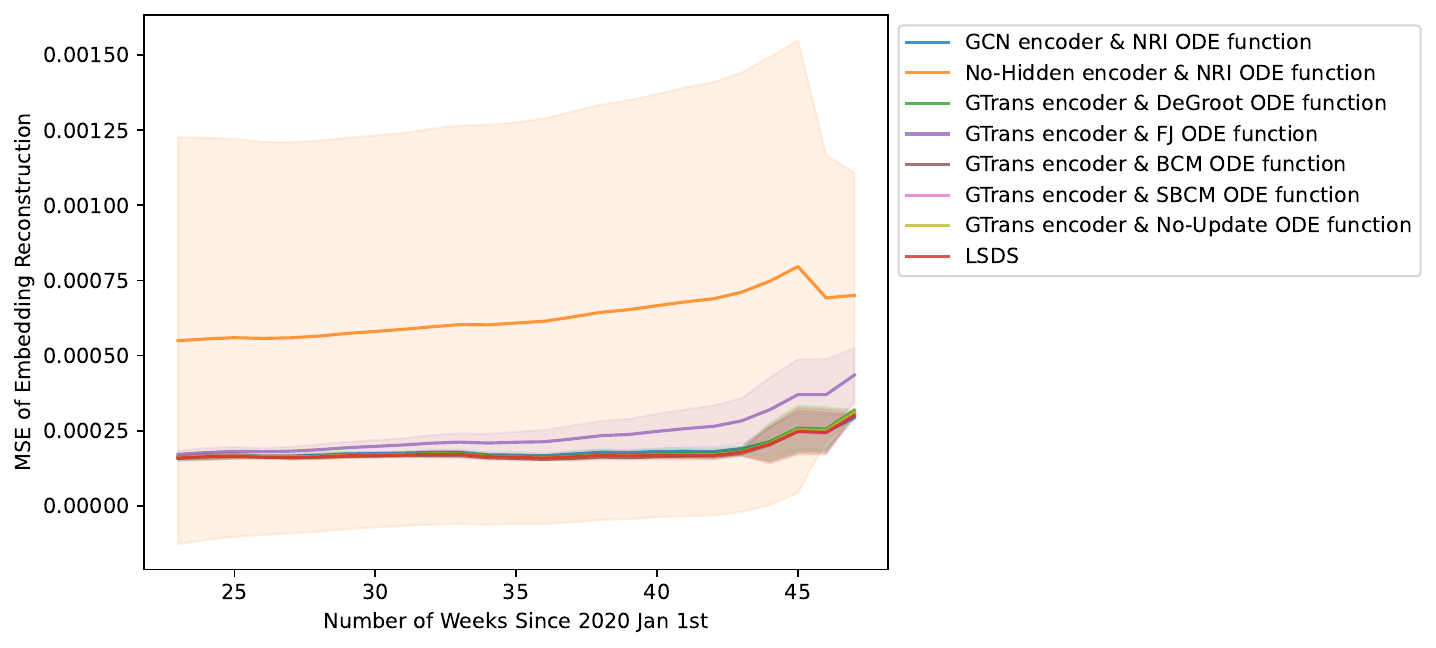}
    \caption{Errors of the text-embedding reconstruction task for different prediction lengths. The encoder uses the first half of the observation sequence, while the second half is used for prediction. The solid line represents the mean values, while the shaded areas indicate the standard deviation.}
    \label{fig:future_polarity_reconstruction}
\end{figure}

From Figures~\ref{fig:future_interaction}, \ref{fig:future_polarity_pred}, \ref{fig:future_polarity_classification} and \ref{fig:future_polarity_reconstruction}, we found our \textbf{\socialdynamical} model always performing better than the other variants in the longer term. When predicting the future behaviors of accounts, errors might accumulate, so it is typically harder for a model to predict the longer-term outcomes.
The high performance of \textbf{\socialdynamical} in long-term prediction suggests that the expressiveness of neural networks enhances the modeling of opinions. By learning from real-world data, it demonstrates strong predictive capabilities over extended time horizons.

Similar to our findings in Section~\ref{sec:performance}, the interaction-prediction task remains the most challenging and serves as the best benchmark for evaluating model performance, as it more clearly differentiates their effectiveness.
In Figure~\ref{fig:future_interaction}, the default \textbf{\socialdynamical} model, equipped with a temporal-graph encoder and NRI ODE function, achieves the best long-term performance, followed by its variant with the BCM (HK) ODE function, which also outperforms most other variants.
However, the model with the \textsc{No-Update} ODE function also achieves high performance in the long term, further highlighting a limitation of the Twitter data set--user behaviors on Twitter exhibit relatively little fluctuation over time.

Aside from the interaction prediction task, the polarity prediction task is the second most effective in distinguishing model performance. From Figure~\ref{fig:future_polarity_pred}, we find that \textbf{\socialdynamical} achieves the best performance in predicting long-term political polarity scores, while the model with the BCM ODE function also performs reasonably well. 
Another interesting finding in Figure~\ref{fig:future_polarity_pred} is that the model with the GCN encoder and NRI ODE function also performs steadily over time, accurately predicting further into the future. This suggests that a flexible ODE function, which captures the updating rules of a social dynamical system as accurately as possible, is crucial for maintaining prediction accuracy over longer time horizons.

Figures~\ref{fig:future_polarity_classification} and \ref{fig:future_polarity_reconstruction} reveal that, in both the polarity classification and text-embedding reconstruction tasks, selecting a suitable encoder is the most important factor in achieving good performance, both in the short term and the long term.

\subsection{Predicting Different Types of Interactions}\label{sec:performance_relations}

\begin{figure}[!htbp]
    \centering
    \includegraphics[width=0.95\linewidth]{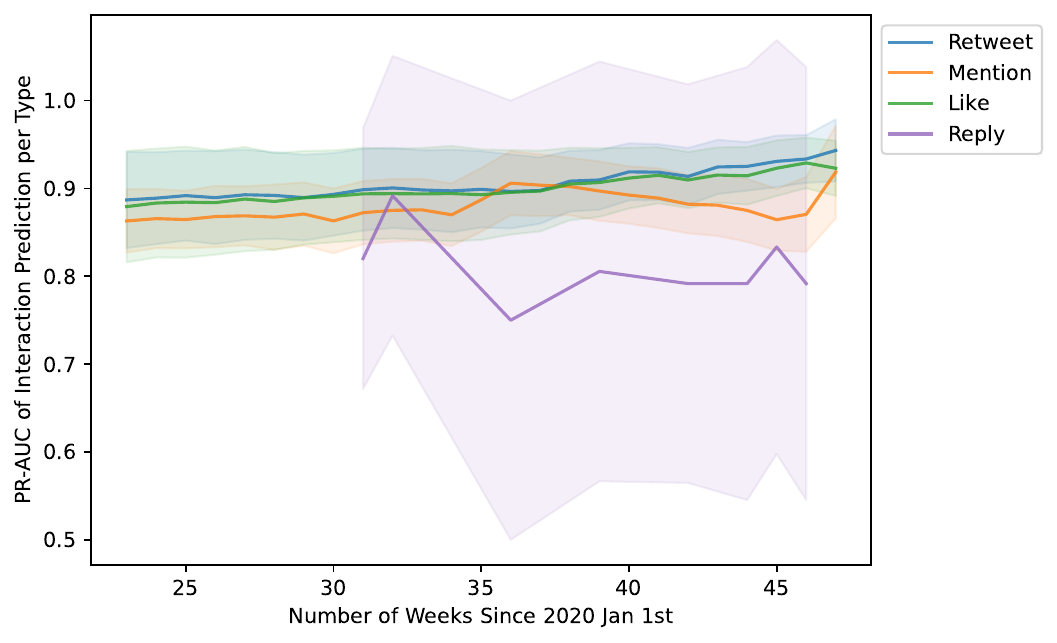}
    \caption{The performance of our model (temporal-graph encoder with NRI ODE function) on the interaction-prediction task for various prediction lengths. Performances are computed separately for different types of interactions. The encoder utilizes the first half of the observation sequence, while the second half is used for prediction. The solid line represents the mean values, while the shaded areas indicate the standard deviation.}
    \label{fig:predict_relations_performance}
\end{figure}

We further investigate the performance of \textbf{\socialdynamical} on the interaction-prediction task and present its performance across different interaction types in Figure~\ref{fig:predict_relations_performance}.

We found that retweet and like interactions are the easiest to predict among all four types. Mention interactions are also relatively straightforward to predict. However, reply interactions, being relatively rare (as shown in Figure~\ref{fig:dataset_interactions}), exhibit low and unstable prediction performance, with a high variance. Predicting reply interactions is the most challenging aspect of the task.

\subsection{Training}\label{sec:dynamic_sys_training}

It is very tricky to train our \textbf{\socialdynamical} model. Because the two parts of the objective in Equation~(\ref{eq::objective}) can conflict with each other, and also because there may exist many local optima point of the decoder loss $\mathcal{L}_{\mathrm{dec}}$, the training performance can easily drop in the middle.
We gradually increase the KL coefficient $\lambda$ (see Section~\ref{sec:dynamic_decoder} and Equation~(\ref{eq:kl_coef})) as the training proceeds.
We compute the KL coefficient $\lambda$ using the formula:
\begin{equation}\label{eq:kl_coef}
    \lambda = \begin{cases}
        0 \,, & (\mathrm{epoch} \leq 10 \text{ or } \mathrm{iter} \leq 10) \\
        \lambda_0 * (1 - 0.99\,^{\mathrm{epoch} - 10}) \,, & (\mathrm{epoch} > 10 \text{ and } \mathrm{iter} > 10)\,,
    \end{cases}
\end{equation}
where $\lambda_0$ is a hyper-parameter that we often set to $0.005$ by default, and we use ``$\mathrm{iter}$'' to name the current number of iterations in a single ``$\mathrm{epoch}$''.
In this approach, the weight of the KL-Divergence objective $\mathcal{L}_{\mathrm{reg}}$ progressively increases from $0$ in each epoch (except the first $10$ epoch where $\lambda$ is set to constant $0$), providing the model greater opportunity to escape local optima, aided by the higher importance of the decoder loss term $\mathcal{L}_{\mathrm{dec}}$ when we have a smaller $\lambda_0$.

\begin{figure}[!htbp]
    \centering
    \includegraphics[width=0.95\linewidth]{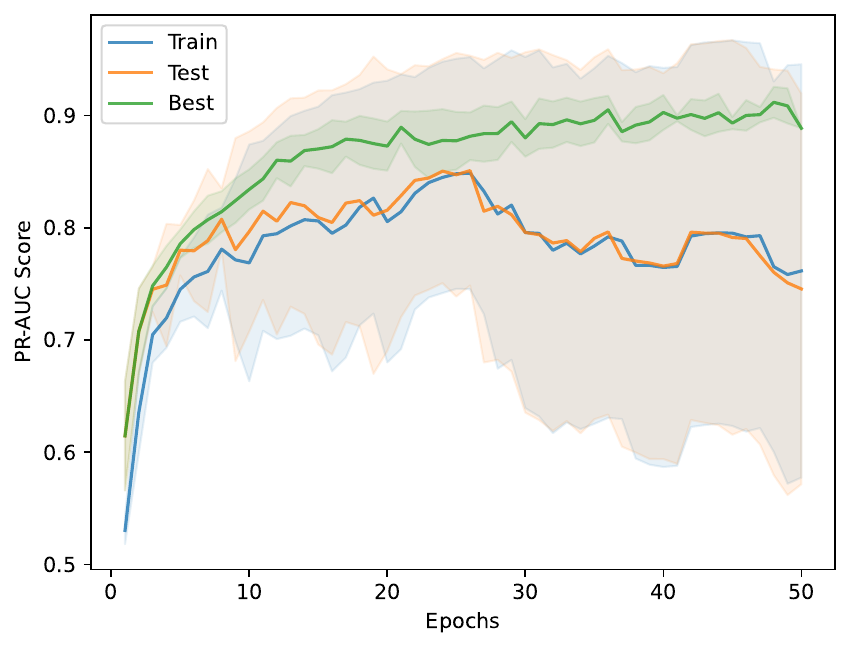} 
    \caption{The training, validation, and testing performance of our model (temporal-graph encoder with NRI ODE function) on the interaction-prediction task across different epochs. The solid line represents the mean values, while the shaded areas indicate the standard deviation.}
    \label{fig:loss}
\end{figure}

In Figure~\ref{fig:loss}, we present the mean values and standard deviations of the performance of our \textbf{\socialdynamical} model (with temporal-graph encoder and NRI ODE function) on the interaction prediction task. We separately plot the training performance, test performance, and the best observed performance (recorded from the first epoch) at each epoch, using the mean values across all five runs. 
Despite our efforts to stabilize the model’s performance, both the training and testing outcomes exhibit significant fluctuations. This underscores the inherent challenges in training a graph ODE framework like \textbf{\socialdynamical}.

\subsection{Model Efficiency}

Our model has reasonable time and memory efficiency both in theory and in practice.
We ran our experiments on a Linux server with the Ubuntu $20.04$ operating system, with a $16G$ RTX A4000 GPU, $32GB$ memory, and $12$ virtual CPUs built on Intel(R) Xeon(R) Gold 5320 CPUs of $2.20G$Hz clock speed.

We use the standard adjoint method for the neural-ODE solver~\cite{zhuang2020adaptive}, so the time complexity of the ODE solver is $\mathcal{O}(N_z N_f (N_t + N_r))$, where $N_z$ is the size of each state $\mathbf{z}_i^t$, $N_f$ is the number of hidden layers (i.e., $L_\mathrm{ODE}$) in the neural-ODE function, and $N_t$ is the number of function evaluations in the forward pass (one can estimate it roughly as the number of time points at which we make predictions), the quantity $N_r$ is the number of evaluations in the backward pass~\cite{zhuang2021mali}. 
To give an example of run time with one of our tasks, running our \textbf{\socialdynamical} model for the temporal-edge prediction task on the \textbf{T20D-Politicians} data set takes approximately $20$--$30$ seconds per epoch.

The number of parameters depends on the choice of the encoder, the ODE function, and the decoder. 

%% file: table-performance.tex
\begin{sidewaystable}
\centering
\caption{Performance on our downstream tasks. We show mean values of results from $5$ different random seeds. The ($\downarrow$) sign indicates the lower the better, and the ($\uparrow$) indicates the higher the better. We show the \textbf{best} results of each column in bold.
}
{
\begin{tabular}{clllllllll}
\hline
\multicolumn{1}{c}{\multirow{2}{*}{}} & 
\multicolumn{2}{c}{\textbf{Interaction-Prediction}} & \multicolumn{2}{c}{\textbf{Polarity Prediction}} & \multicolumn{2}{c}{\textbf{Polarity Classification}} & \multicolumn{3}{c}{\textbf{Text-Embedding Reconstruction}} \\ \cmidrule{2-3}\cmidrule{4-5}\cmidrule{6-7}\cmidrule{8-10}
\multicolumn{1}{c}{\textbf{Model Variants}} & \multicolumn{1}{c}{ROC-AUC ($\uparrow$)} & \multicolumn{1}{c}{PR-AUC ($\uparrow$)} & 
\multicolumn{1}{c}{MSE ($\downarrow$)} & \multicolumn{1}{c}{MAPE ($\downarrow$)} & 
\multicolumn{1}{c}{Accuracy ($\uparrow$)} & \multicolumn{1}{c}{$F_1$-score ($\uparrow$)} & 
\multicolumn{1}{c}{MSE ($\downarrow$)} & \multicolumn{1}{c}{MAPE ($\downarrow$)} & \multicolumn{1}{c}{$R^2$-score ($\uparrow$)} \\ \hline

\multicolumn{10}{c}{\textbf{\socialdynamical}  with other encoder variants} \\ \hline

\multicolumn{1}{c|}{\textsc{GCN}} & 
\multicolumn{1}{c|}{ 0.8087 } & \multicolumn{1}{c|}{ 0.7881 } & 
\multicolumn{1}{c|}{ 0.2225 } & \multicolumn{1}{c|}{ 1.9510 } & 
\multicolumn{1}{c|}{ 0.8635 } & \multicolumn{1}{c|}{ 0.8633 } & 
\multicolumn{1}{c|}{ 0.0108 } & \multicolumn{1}{c|}{ 5.4442 } & \multicolumn{1}{c}{ 0.3653 }  \\
\multicolumn{1}{c|}{\textsc{No-Hidden}} & 
\multicolumn{1}{c|}{ 0.8164 } & \multicolumn{1}{c|}{ 0.8001 } &
\multicolumn{1}{c|}{ 0.2674 } & \multicolumn{1}{c|}{ 3.2792 } & 
\multicolumn{1}{c|}{ 0.8351 } & \multicolumn{1}{c|}{ 0.8349 } & 
\multicolumn{1}{c|}{ 0.0316 } & \multicolumn{1}{c|}{ 13.3537 } & \multicolumn{1}{c}{ -0.7352 }  \\ \hline

\multicolumn{10}{c}{\textbf{\socialdynamical}  with other ODE function variants} \\ \hline

\multicolumn{1}{c|}{\textsc{DeGroot}} & 
\multicolumn{1}{c|}{ 0.9064 } & \multicolumn{1}{c|}{ 0.8908 } & 
\multicolumn{1}{c|}{ \textbf{0.1731} } & \multicolumn{1}{c|}{ 3.0134 } & 
\multicolumn{1}{c|}{ 0.8939 } & \multicolumn{1}{c|}{ 0.8938 } & 
\multicolumn{1}{c|}{ \textbf{0.0094} } & \multicolumn{1}{c|}{ 5.0998 } & \multicolumn{1}{c}{ \textbf{0.4438} } \\

\multicolumn{1}{c|}{\textsc{FJ}} & 
\multicolumn{1}{c|}{ 0.8984 } & \multicolumn{1}{c|}{ 0.8784 } & 
\multicolumn{1}{c|}{ 0.1792 } & \multicolumn{1}{c|}{ 3.2393 } & 
\multicolumn{1}{c|}{ \textbf{0.8946} } & \multicolumn{1}{c|}{ \textbf{0.8946} } & 
\multicolumn{1}{c|}{ 0.0095 } & \multicolumn{1}{c|}{ 5.1118 } & \multicolumn{1}{c}{ 0.4405 }  \\

\multicolumn{1}{c|}{\textsc{BCM} (\textsc{HK})} & 
\multicolumn{1}{c|}{ 0.9103 } & \multicolumn{1}{c|}{ 0.8958 } & 
\multicolumn{1}{c|}{ \textbf{0.1731} } & \multicolumn{1}{c|}{ 2.8501 } & 
\multicolumn{1}{c|}{ 0.8940 } & \multicolumn{1}{c|}{ 0.8940 } & 
\multicolumn{1}{c|}{ \textbf{0.0094} } & \multicolumn{1}{c|}{ \textbf{5.0465} } & \multicolumn{1}{c}{ 0.4432 }  \\

\multicolumn{1}{c|}{\textsc{SBCM}} & 
\multicolumn{1}{c|}{ 0.8954 } & \multicolumn{1}{c|}{ 0.8785 } & 
\multicolumn{1}{c|}{ 0.1743 } & \multicolumn{1}{c|}{ 3.0387 } & 
\multicolumn{1}{c|}{ 0.8940 } & \multicolumn{1}{c|}{ 0.8940 } & 
\multicolumn{1}{c|}{ 0.0100 } & \multicolumn{1}{c|}{ 5.3295 } & \multicolumn{1}{c}{ 0.4079 }  \\

\multicolumn{1}{c|}{\textsc{No-Update}} & 
\multicolumn{1}{c|}{ 0.8950 } & \multicolumn{1}{c|}{ 0.8737 } & 
\multicolumn{1}{c|}{ 0.1741 } & \multicolumn{1}{c|}{ 2.9019 } & 
\multicolumn{1}{c|}{ 0.8935 } & \multicolumn{1}{c|}{ 0.8935 } & 
\multicolumn{1}{c|}{ 0.0099 } & \multicolumn{1}{c|}{ 5.4087 } & \multicolumn{1}{c}{ 0.4169 }  \\ \hline

\multicolumn{1}{c|}{\textbf{\socialdynamical} } & 
\multicolumn{1}{c|}{ \textbf{0.9119} } & \multicolumn{1}{c|}{ \textbf{0.8997} } & 
\multicolumn{1}{c|}{ 0.1770 } & \multicolumn{1}{c|}{ \textbf{1.7894} } & 
\multicolumn{1}{c|}{ 0.8925 } & \multicolumn{1}{c|}{ 0.8925 } & 
\multicolumn{1}{c|}{ 0.0097 } & \multicolumn{1}{c|}{ 5.1769 } & \multicolumn{1}{c}{ 0.4281 }  \\ \hline
\end{tabular}
} 
\label{tab:performance}
\end{sidewaystable}


%% file: 6-conclusion.tex
\section{Conclusion and Future Work}\label{sec:sds_sonclusion}

We introduced {\fullname} (\textbf{\socialdynamical}), a framework for modeling multi-agent dynamical systems on real-world social networks. Using Twitter data, we evaluated its performance on challenging tasks such as temporal interaction prediction and political polarity prediction. Our results demonstrate that \textbf{\socialdynamical} effectively captures the evolving dynamics of opinions and interactions. Moreover, its modular design allows seamless integration of different encoders, decoders, and ODE functions, making it adaptable to diverse downstream tasks and data sets.


There are several exciting directions for future research:
\begin{enumerate}
    \item \textbf{Integrating External Information Sources:} 
    Our current model only leverages social-media data. However, real-world opinion formation is influenced by various external factors, including news, television, and offline discussions. Extending \textbf{\socialdynamical} to incorporate multi-modal data from these sources is a challenging yet promising avenue.
    \item \textbf{Modeling Adaptive Networks:}
    In reality, social networks evolve—nodes (users) and edges (relationships) appear, disappear, and change over time. Generalizing \textbf{\socialdynamical} to adaptive networks~\cite{berner2023adaptive} would enable the study of more dynamic social structures.
    \item \textbf{Exploring Alternative Neural-ODE Functions:}
    While the NRI model provided a robust ODE function for most tasks, it was not explicitly designed for social dynamical systems. Investigating alternative graph neural networks (GNNs) could improve the modeling of opinion dynamics and other social behaviors.
\end{enumerate}

Applying \textbf{\socialdynamical} to real-world data sets requires careful ethical consideration. As with any neural network, our model may inherit biases present in training data, such as gender bias. Furthermore, despite achieving strong performance on specific tasks, predictions made by \textbf{\socialdynamical} should never be interpreted as absolute truths about individuals. Ensuring responsible use and continuous assessment of potential biases will be crucial in future applications.

%% file: 7-declarations.tex

\section*{Declarations}


\subsection*{List of Abbreviations}


\begin{table}[h]
    \caption{List of Abbreviations}
    \centering
    \begin{tabular}{c|c}
    \toprule
        \textbf{Abbreviation} & \textbf{Corresponding Term} \\  \midrule
        LSDS & {L}atent {S}ocial {D}ynamical-{S}ystem \\
        ODE & Ordinary Differential Equation \\
        GraphODE & GNN-based Neural Ordinary Differential Equation \\
        Neural-ODE & Neural Ordinary Differential Equation \\
        GNN & Graph Neural Network \\
        NN & Neural Network \\
        DNRL & Dynamic-Node-Representation Learning \\ 
        TSA & Temporal Self-Attention \\
        PEM & polarity-aware embedding model \\
        LG-ODE & Latent Graph Ordinary Differential Equation \\
        VAE & Variational Autoencoder \\ 
        T20D-PolA & Twitter 2020 Dynamical data of Political-related Accounts \\
        
    \bottomrule 
    \end{tabular}
    \label{tab:list_abbr}
\end{table}

In Table~\ref{tab:list_abbr}, we list the most important abbreviations in our paper.


\subsection*{Ethics Approval and Consent to Participate}

Not applicable.


\subsection*{Consent for Publication}

Not applicable.


\subsection*{Availability of Data and Materials}


We have submitted our code and a subset of our data as supplementary materials with our paper.
We promise to share our data and code.
Once our paper is published, we will release our full data set on a public repository.


\subsection*{Competing interests}


Not applicable.


\subsection*{Funding}




This research was supported by the National Science Foundation (through grants III-1705169, NSF 1937599, NSF 2119643, and 1922952), an Okawa Foundation Grant, Amazon Research Awards, Cisco research grant USA000EP280889, Picsart Gifts, and Snapchat Gifts.


\subsection*{Authors' Contributions}


ZX, ZH, MAP, and YS conceived and conceptualized the study. 
ZX, XW, YQ, and ZH performed the analysis and wrote the initial draft of the paper.
ZX, MAP, and YS reviewed and extensively edited the manuscript, determined what additional analysis was necessary, and produced the final version of the manuscript.
All authors read and approved the final manuscript.


\subsection*{Acknowledgements}


We thank Fang Sun and Xiao Luo for helpful discussions.